\documentclass[final,fleqn,5p]{elsarticle}
\usepackage{graphicx}
\usepackage{amssymb}
\usepackage{amsthm}
\usepackage{amsmath}
\usepackage{lineno}
\usepackage{multirow}
\usepackage{caption}
\usepackage{mathrsfs}
\usepackage{sectsty}
\usepackage{epsfig}
\usepackage{pmat}
\theoremstyle{definition}

\newtheorem{algorithm}{Algorithm}

\usepackage[version=3]{mhchem}
\usepackage[table]{xcolor}

\usepackage{natbib}
  
\DeclareMathOperator*{\argmax}{arg\,max}

\usepackage{savesym}
\savesymbol{AND}

\usepackage{algorithmic}
\usepackage{algorithm}

\sectionfont{\normalsize}
\subsectionfont{\normalsize \it} 
\usepackage{enumitem}
\setlist{nolistsep,leftmargin=*}

\journal{ArXiv}
\begin{document}
\begin{frontmatter}

\title{{Deep Reinforcement Learning for Process Control: A Primer for Beginners}}
\author[UBC]{Steven Spielberg}
\author[UBC]{Aditya Tulsyan}
\author[UBC1]{Nathan P. Lawrence}
\author[UBC1]{Philip D Loewen}
\author[UBC]{R. Bhushan Gopaluni\corref{cor1}}
\address[UBC]{Department of Chemical and Biological Engineering, University of British Columbia, Vancouver, BC V6T 1Z3, Canada.}
\address[UBC1]{Department of Mathematics, University of British Columbia, Vancouver, BC V6T 1Z2, Canada.}
\cortext[cor1]{\small{Corresponding author. Email: bhushan.gopaluni@ubc.ca}}
\cortext[cor2]{\small{Please cite this paper as: Toward self-driving processes: A deep reinforcement learning approach to control. AIChE Journal, Vol. 65, No. 10, 2019.}}

\begin{abstract}
Advanced model-based controllers are well established in process industries.
However, such controllers require regular maintenance to maintain acceptable performance. 
It is a common practice to monitor controller performance continuously
and to initiate a remedial model re-identification procedure in the event of performance degradation. 
Such procedures are typically complicated and resource-intensive, 
and they often cause costly interruptions to normal operations.  
In this paper, we exploit recent developments in reinforcement learning and deep learning to develop a novel adaptive, model-free controller for general discrete-time processes. 
The DRL controller we propose is a data-based controller that learns the control policy in real time by merely interacting with the process. 
The effectiveness and benefits of the DRL controller are demonstrated through many simulations. 
\end{abstract}


\end{frontmatter}

\section{{Introduction}}
\label{sec:Introduction}
Industrial process control is a large and diverse field;
its broad range of applications calls for a correspondingly wide range of controllers---including single and multi-loop PID controllers, 
model predictive controllers (MPCs), and a variety of nonlinear controllers. 
Many deployed controllers achieve robustness at the expense of performance.  
The overall performance of a controlled process depends on the characteristics
of the process itself, the controller's overall architecture, and the tuning parameters that are employed. 
Even if a controller is well-tuned at the time of installation,
drift in process characteristics or deliberate set-point changes 
can cause performance to deteriorate over time \cite{tulsyan2018machine,tulsyan2018advances,tulsyan2019industrial}.
Maintaining system performance over the long-term is essential.
Unfortunately, it is typically also both complicated and expensive.

Most modern industrial controllers are model-based, 
so good performance calls for a high-quality process model. 
It is standard practice to continuously monitor system performance and initiate a remedial model re-identification exercise in the event of performance degradation. 
Model re-identification can require two weeks or more \cite{kano2009state}, 
and typically involves the injection of external excitations \cite{tulsyan2013bayesian}, 
which introduce an expensive interruption to the normal operation of the process. 
Re-identification is particularly complicated for multi-variable processes,
which require a model for every input-output combination.  

Most classical controllers in industry are linear and non-adaptive. 
While extensive work has been done in nonlinear adaptive control~\cite{seborg1986adaptive, krstic1995nonlinear}, 
it has not yet established a significant position in the process industry, 
beyond several niche applications \cite{mailleret2004nonlinear, guan2008adaptive}. 
The difficulty of learning (or estimating) reliable multi-variable models in an online fashion is partly responsible for this situation.  
Other contributing factors include the presence of hidden states, 
process dimensionality, and in some cases computational complexity.

Given the limitations of existing industrial controllers, 
we seek a new design that can learn the control policy for discrete-time nonlinear stochastic processes in real time,
in a model-free and adaptive environment.  
This paper continues our recent investigations \cite{spielberg2017deep} on the same topic. 
The idea of RL has been around for several decades; 
however, its application to process control has been somewhat recent. 
Next, we provide a short introduction to RL and highlight RL-based approaches for control problems. 

\subsection{Reinforcement Learning  and Process Control}
Reinforcement Learning (RL) is an active area of research in artificial intelligence.
It originated in computer science and operations research 
to solve complex sequential decision-making problems  \cite{sutton1998reinforcement, bertsekas1995neuro, bertsekas2005dynamic, powell2007approximate}.  
The RL framework comprises an {agent} (e.g., a controller) 
interacting with a stochastic {environment} (e.g., a plant) modelled as a Markov decision process (MDP). 
The goal in an RL problem is to find a policy 
that is optimal in certain sense \cite{sugiyama2015statistical}.  

{%
Over the last three decades, 
several methods, including {dynamic programming} (DP), 
{Monte Carlo} (MC)  and {temporal-difference learning} (TD) 
have been proposed to solve the RL problem \cite{sutton1998reinforcement}}. 
Most of these compute the optimal policy using {policy iteration}. 
This is an iterative approach in which every step involves both 
{policy estimation} and {policy improvement}. 
The policy estimation step aims at making the value function 
consistent with the current policy; 
the policy improvement step makes the policy {greedy} 
with respect to the current estimate of the value function. 
Alternating between the two steps produces a sequence of value function estimates and sub-optimal policies
which, in the limit, converge to the optimal value function and the optimal policy, respectively \cite{sutton1998reinforcement}.  

The choice of a solution strategy for an RL problem is primarily driven by the assumptions on the environment and the agent. 
For example, under the perfect model assumption for the environment, 
classical dynamic programming methods for Markov Decision Problems with finite state and action spaces)
provide a closed-form solution to the optimal value function, 
and are known to converge to the optimal policy in polynomial time  \cite{bertsekas2005dynamic, bellman2013dynamic}.  
Despite their strong convergence properties, however,
classical DP methods have limited practical applications because of their stringent requirement for a perfect model of the environment, which is seldom available in practice. 

Monte Carlo algorithms belong to a class of approximate RL methods 
that can be used to estimate the value function using {experiences} 
(i.e., sample sequences of states, actions, and rewards accumulated through
the agent's interaction with an environment). 
MC algorithms offer several advantages over DP. 
First, MC methods allow an agent to learn the optimal behaviour directly by interacting with the environment,
thereby eliminating the need for an exact model. 
Second, MC methods can be focused, to estimate value functions for a small subset of the states of interest rather than evaluating them for the entire state space, as with DP methods. 
This significantly reduces the computational burden,
since value function estimates for less relevant states need not be updated. 
Third, MC methods may be less sensitive to the violations of the Markov property of the system. 
Despite the advantages mentioned above, MC methods are difficult to implement in real time, 
as the value function estimates can only be updated at the end of an experiment. 
{
Further,  MC methods are also known to exhibit slow convergence as they do not {bootstrap}, i.e., they do not update their value function from other value estimates \cite{sutton1998reinforcement}.} 
The efficacy of MC methods remain unsettled and is a subject of  research \cite{sutton1998reinforcement}.  

TD learning is another class of approximate methods for solving RL problems that combine ideas from DP and MC. 
Like MC algorithms, TD methods can learn directly from raw experiences without requiring a model; 
and like DP, TD methods update the value function in real time without having to wait until the end of the experiment. 
For a detailed exposition on RL solutions, the reader is referred to \cite{sutton1998reinforcement} and the references therein.


Reinforcement Learning has achieved remarkable success in robotics \cite{lehnert2015policy, lillicrap2015continuous}, 
computer games \cite{mnih2013playing,mnih2015human}, online advertising \cite{pednault2002sequential}, 
and board games \cite{tesauro1995temporal, silver2016mastering};
however, its adaptation to  process control has been limited \textcolor{black}{(see  \citet{badgwell2018reinforcement} for recent survey of RL methods in process control)} ---even though many optimal scheduling and control problems 
can be formulated as MDPs \cite{lee2006approximate}. 
This is primarily due to lack of efficient RL algorithms to deal with infinite MDPs 
(i.e., MDPs with continuous state and action spaces) 
that define most modern control systems. 
\textcolor{black}{While existing RL methods apply to infinite MDPs, exact solutions are possible only in special cases, 
such as in linear quadratic (Gaussian) control problem, where DP provides a closed-form solution \cite{lewis2012reinforcement, morinelly2016dual}.} 
It is plausible to discretize infinite MDPs and use DP to estimate the value function; however, this leads to an exponential growth in the computational complexity with respect to the states and actions, which is referred to as {curse of dimensionality} \cite{bertsekas2005dynamic}. 
The computational and storage requirements for discretization methods applied to most problems of practical interest in process control remain unwieldy even with today’s computing hardware \cite{lee2006choice}. 

\color{black} 
The first successful implementation of RL in process control appeared in the series of papers published in the early 2000s 
\cite{lee2006approximate, lee2006choice, lee2001neuro, kaisare2003simulation, lee2004simulation, lee2005approximate}, 
where the authors proposed approximate dynamic programming (ADP) for optimal control of discrete-time nonlinear systems. 
The idea of ADP is rooted in the formalism of DP but uses simulations and function approximators (FAs) 
to alleviate the curse of dimensionality.  
Other RL methods based on heuristic dynamic programming (HDP) \cite{al2008discrete}, direct HDP \cite{si2000online}, 
dual heuristic programming \cite{heydari2013finite}, and globalized DHP \cite{wang2012optimal} 
have also been proposed for optimal control of discrete-time nonlinear systems. 
RL methods have also been proposed for optimal control of continuous-time nonlinear systems 
\cite{vrabie2009neural, vamvoudakis2010online, liu2014decentralized}. 
However, unlike discrete-time systems, controlling continuous-time systems with RL has proven to be considerably more difficult and fewer results are available \cite{lewis2012reinforcement}.  
While the contributions mentioned above establish the feasibility and adaptability of RL in controlling discrete-time
 and continuous-time nonlinear processes, most of these methods assume complete or partial access to process models 
\cite{lee2006approximate, lee2006choice, lee2001neuro, kaisare2003simulation, lee2004simulation, vamvoudakis2010online, liu2014decentralized}. 
This limits existing RL methods to processes for which high-accuracy models are either available or can be derived through system identification. 

Recently, several data-based approaches have been proposed to address the limitations of model-based RL in control.  
In \citet{lee2005approximate, mu2017novel}, a data-based learning algorithm was proposed to derive an improved control policy for discrete-time nonlinear systems using ADP with an identified process model, 
as opposed to an exact model. 
Similarly,  \citet{lee2005approximate} proposed a Q-learning algorithm to learn an improved control policy in a model-free manner using only input-output data.  
While these methods remove the requirement for having an exact model (as in RL), 
they still present several issues. 
For example, the learning method proposed in \citet{lee2005approximate, mu2017novel} is still based on ADP, 
so its performance relies on the accuracy of the identified model. 
For complex, nonlinear, stochastic systems, identifying a reliable process model may be nontrivial as it often requires running multiple carefully designed experiments. 
Similarly, the policy derived from Q-learning in \citet{lee2005approximate} 
may converge to a sub-optimal policy as it avoids adequate exploration of the state and action spaces. 
Further, calculating the policy with Q-learning for infinite MDPs requires solving a non-convex optimization problem over the continuous action space at each sampling time, 
which may render it unsuitable for online deployment for processes with small time constants or modest computational resources.  
Note that data-based RL methods have also been proposed for continuous-time nonlinear systems \cite{luo2014data, wang2016data}.
Most of these data-based methods approximate the solution to the Hamilton-Jacobi-Bellman (HJB) equation  derived for a class of continuous-time affine nonlinear systems using policy iteration.
\textcolor{black}{For more information on RL-based optimal control of continuous-time nonlinear systems,  the reader is referred to \citet{luo2014data, wang2016data, tang2018distributed} and the references therein. }

The promise of RL to deliver a real-time, 
self-learning controller in a model-free and adaptive environment has long motivated the process systems community to explore novel approaches to apply RL in process control applications. 
While the plethora of studies from over last two decades has provided significant insights to help connect the RL paradigm with process control, 
they also highlight the nontrivial nature of this connection and the limitations of existing RL methods in control applications. 
Motivated by recent developments in the area of deep reinforcement learning (DRL), 
we revisit the problem of RL-based process control and explore the feasibility of 
using DRL to bring these goals one step closer.

\begin{figure}[t]
	\centering
	\includegraphics[width=0.3\textwidth]{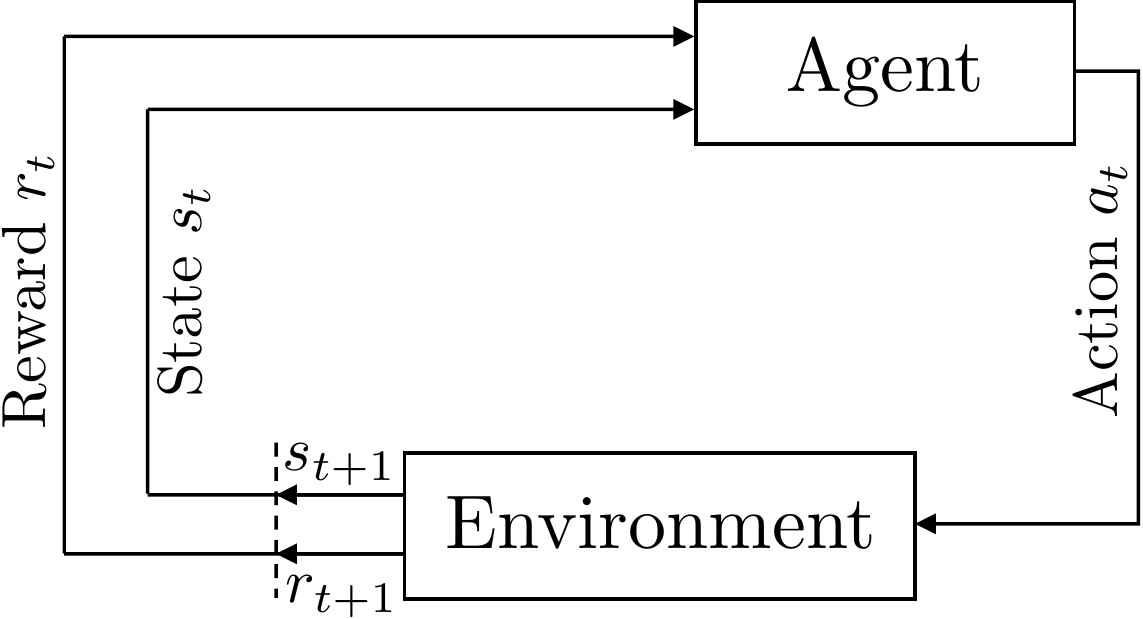}
	\caption{A schematic of the agent-environment interactions in a standard RL problem.}
	\label{fig:rl_block_diagram}
\end{figure}

\subsection{Deep Reinforcement Learning (DRL)}
The recent resurgence of interest in RL results from the successful combination 
of RL with deep learning that allows for effective generalization of RL to MDPs
 with continuous state spaces. 
The {Deep-Q-Network} (DQN) proposed recently in \citet{mnih2013playing} 
combines deep learning for sensory processing \cite{krizhevsky2012imagenet} 
with RL to achieve human-level performance on many \texttt{Atari} video games. 
Using pixels as inputs, simply through interactions, 
the DQN can learn in real time the optimal strategy to play \texttt{Atari}.  
This was made possible using a deep neural network function approximator (FA)
to estimate the action-value function over the continuous state space, 
which was then maximized over the action space to find the optimal policy. 
Before DQN, learning the action-value function using FAs was widely considered difficult and unstable. 
Two innovations account for this latest gain in stability and robustness: 
(a) the network is trained off-policy with samples from a replay buffer 
to minimize correlations between samples, and 
(b) the network is trained with a target Q-network to give consistent targets during temporal difference backups.  
{
In this paper, we propose an off-policy actor-critic algorithm, 
referred to as a DRL controller,
for controlling of discrete-time nonlinear processes. 
The proposed DRL controller is a model-free controller designed based on TD learning.}  
As a data-based controller, 
the DRL controller uses two independent deep neural networks to generalize the actor and critic to continuous state and action spaces. 
The DRL controller is based on the deterministic policy gradient (DPG) algorithm proposed by 
\citet{silver2014deterministic} that combines an actor-critic method with insights from DQN. 
The DRL controller uses ideas similar to those in \citet{lillicrap2015continuous}, 
modified to make learning suitable for process control applications. 
Several examples of different complexities are presented to demonstrate the efficacy of the DRL controller in set-point tracking problems.

The rest of the paper is organized as follows:  in Section \ref{sec:RLProblem}, we introduce the basics of MDP  and derive a  control policy based on value functions.  Motivated by the intractability of optimal policy calculations for infinite MDPs,  we introduce Q-learning in Section \ref{sec:QLearning} 
for solving RL problem over continuous state space.  A policy gradient algorithm is discussed in Section \ref{sec:DPG} 
to extend the RL solution to continuous action space. 
Combining the developments in Sections \ref{sec:QLearning} and \ref{sec:DPG}, 
a novel actor-critic algorithm is discussed in Section \ref{sec:ActorCritic}. 
In Section \ref{sec:RLProcessControl}, 
a DRL framework is proposed for data-based control of discrete-time nonlinear processes. 
The efficacy of a DRL controller is demonstrated  on several  examples in 
Section~\ref{sec:Simulation}. 
Finally, Section \ref{sec:DRLVsMPC} compares a DRL controller to an MPC. 

This paper follows a tutorial-style presentation to assist readers unfamiliar with the theory of RL. 
The material is systematically introduced to highlight the challenges with existing RL solutions 
in process control applications and to motivate the development of the proposed DRL controller. 
The background material presented here is only introductory, 
and readers are encouraged to refer to the cited references for a detailed exposition on these topics.

\section{The Reinforcement Learning (RL) Problem}
\label{sec:RLProblem}
The RL framework consists of a learning agent (e.g., a controller) interacting with 
a stochastic environment (e.g., a plant or process), denoted by $\mathcal{E}$, in discrete time steps. 
The objective in an RL problem is to identify a policy (or control actions) to maximize 
the expected cumulative reward the agent receives in the long run \cite{sutton1998reinforcement, sugiyama2015statistical}. 
The RL problem is a sequential decision-making problem, 
in which the agent incrementally learns how to optimally interact with the environment by maximizing the expected reward it receives.  
Intuitively, the agent-environment interaction is to be understood as follows. 
Given a {state space} $\mathcal{S}$ and an {action space} $\mathcal{A}$, 
the agent at time step $t\in\mathbb{N}$ observes some representation of the environment's state $s_t\in\mathcal{S}$ 
and on that basis selects an action $a_t\in\mathcal{A}$. 
One time step later, in part as a consequence of its action, 
the agent finds itself in a new state  $s_{t+1}\in\mathcal{S}$ and receives a scalar reward $r_t\in\mathbb{R}$ 
from the environment indicating how well the agent performed at $t\in\mathbb{N}$. 
This procedure repeats for all $t\in\mathbb{N}$, as in the case of a {continuing task} 
or until the end of an {episode}.  
The agent-environment interactions are illustrated in Figure \ref{fig:rl_block_diagram}.

\subsection{Markov Decision Process (MDP)}

A Markov Decision Process (MDP) consists of the following: 
a state space $\mathcal{S}$; 
an action space $\mathcal{A}$; 
an initial state distribution $p(s_1)$; 
and a transition distribution $p(s_{t+1}|s_t,a_t)$\footnote{To simplify notation, 
   we drop the random variable in the conditional density and write 
   $p(s_{t+1}|s_t, a_t) = p(s_{t+1}|S_t = s_t, A_t = a_t)$.} 
satisfying the following Markov property 
\begin{align}
\label{eq:MarkovP}
p(s_{t+1}|s_t,a_t,\dots,s_1,a_1)=p(s_{t+1}|s_t,a_t),
\end{align}
for any trajectory $s_1,a_1\dots,s_T,a_T$ generated in the state-action space 
$\mathcal{S}\times\mathcal{A}$ 
(for continuing tasks, $T\rightarrow \infty$, while for episodic tasks, $T\in\mathbb{N}$ is the terminal time); 
and a reward function $r_t\colon\mathcal{S}\times\mathcal{A}\rightarrow\mathbb{R}$. 
In an MDP, the transition function (\ref{eq:MarkovP}) is the likelihood 
of observing a state $s_{t+1}\in\mathcal{S}$ after the agent takes an action 
$a_t\in\mathcal{A}$ in state $s_t\in\mathcal{S}$.
Thus $\int_\mathcal{S}p(s|s_t,a_t)\,ds = 1$ for all $(s_t,a_t)\in\mathcal{S}\times\mathcal{A}$.

A {policy} is used to describe the actions of an agent in the MDP. 
A stochastic policy is denoted by $\pi\colon\mathcal{S} \rightarrow \mathcal{P}(\mathcal{A})$, 
where $\mathcal{P}(\mathcal{A})$ is the set of probability measures on $\mathcal{A}$.
Thus $\pi(a_t|s_t)$ is the probability of taking action $a_t$ when in state $s_t$;
we have $\int_{\mathcal{A}}\pi(a|s_t)\,da=1$ for each $s_t\in\mathcal{S}$.
This general formulation allows any deterministic policy $\mu\colon\mathcal{S}\rightarrow\mathcal{A}$,
through the definition
\begin{align}
\pi(a|s_t)=
\begin{cases}
1\quad \text{if~} a=\mu(s_t),\\
0 \quad \text{otherwise}.
\end{cases}
\end{align}
The agent uses $\pi$ to sequentially interact with the environment 
to generate a sequence of states, actions and rewards in $\mathcal{S}\times\mathcal{A}\times\mathcal{R}$, 
denoted generically as follows $h=(s_1,a_1,r_1,\dots,s_T,a_T,r_T)$, 
where $(H=h)\sim p^\pi(h)$ is an arbitrary history generated under policy $\pi$ 
and distributed according to the probability density function (PDF) $p^\pi(\cdot)$. 
Note that under the Markov assumption of the MDP, the PDF for the history can be decomposed as follows:
\[
p^\pi(h) = p(s_1)\prod_{t=1}^Tp(s_{t+1}|s_t,a_t)\pi(a_t|s_t).
\]
The total reward accumulated by the agent in a task from time $t\in\mathbb{N}$ onward is
\begin{align}
\label{eq:Reward}
R_t(h)=\sum_{k=t}^{\infty} \gamma^{k-t}r_t(s_k,a_k),
\end{align}
Here $\gamma \in [0,1]$ is a (user-specified) discount factor:
a unit reward has the value $1$ right now, $\gamma$ after one time step,
and $\gamma^{\tau}$ after $\tau$ sampling intervals.
If we specify $\gamma=0$, only the immediate reward has any influence;
if we let $\gamma\rightarrow 1$, future rewards are considered more strongly---informally,
the controller becomes `farsighted.'
While (\ref{eq:Reward}) is defined for continuing tasks, i.e., $T\rightarrow\infty$, 
the notation can be applied also for episodic tasks by introducing
a special {absorbing} terminal state that transitions only to itself and generates rewards of $0$.
These conventions are used to simplify the notation and to express close parallels 
between episodic and continuing tasks. 
See \citet{sutton1998reinforcement}.
 
\subsection{Value Function and Optimal Policy}
\label{sec:ValueFunction}
The {state-value function}, 
$V^\pi\colon\mathcal{S}\rightarrow\mathbb{R}$ 
assigns a value to each state, according to the given policy $\pi$. 
In state $s_t$, $V^\pi(s_t)$ is the value of the future reward 
an agent is expected to receive by starting at $s_t\in\mathcal{S}$ 
and following policy $\pi$ thereafter.
In detail,
\begin{align}
\label{eq:VF}
V^\pi(s_t) &= \mathbb{E}_{h\sim p^\pi(\cdot)}[R_t(h)|s_t].
\end{align}
The closely-related {action-value} function
$Q^\pi\colon\mathcal{S}\times\mathcal{A}\rightarrow\mathbb{R}$
decouples the immediate action from the policy $\pi$, 
assuming only that $\pi$ is used for all subsequent steps: 
\begin{align}
\label{eq:AF}
Q^\pi(s_t,a_t) = \mathbb{E}_{h\sim p^\pi(\cdot)}[R_t(h)|s_t,a_t].
\end{align}
The Markov property of the underlying process
gives these two value functions a recurrent structure illustrated below:
\begin{align}
\label{eq:BE_AF}
Q^\pi(s_t, a_t) =&\mathbb{E}_{s_{t+1}\sim p(\cdot|s_t,a_t)}\left[r(s_t, a_t) + \gamma\pi(a_{t+1}|s_{t+1})\right.\nonumber\\
&\left. \times Q^\pi(s_{t+1}, a_{t+1})\right].
\end{align}
Solving an RL problem amounts to finding $\pi^\star$ that outperforms all other policies
across all possible scenarios.
Identifying $\pi^*$ will yield an optimal Q-function, $Q^\star$, such that
\begin{align}
\label{eq:OAF}
Q^\star(s_t,a_t) = \max_{\pi} Q^\pi(s_t,a_t),\qquad (s_t,a_t)\in\mathcal{S}\times\mathcal{A}.
\end{align}
Conversly, knowing the function  $Q^\star$ is enough to recover an optimal policy
by making a ``greedy'' choice of action:
\begin{align}
\label{eq:OP}
\pi^\star(a|s_t) 
=
\begin{cases}
1,\quad \text{if } a = \argmax_{a\in\mathcal{A}} Q^\star(s_t,a),\\
0,\quad \text{otherwise}.\\
\end{cases}
\end{align}
Note that this policy is actually deterministic.
%
While solving the Bellman equation in (\ref{eq:BE_AF}) for the Q-function provides an approach 
to finding an optimal policy in (\ref{eq:OP}), and thus solving the RL problem, this solution is rarely useful in practice. 
This is because the solution relies on two  key assumptions -- (a) the dynamics of the environment is accurately known, i.e., $p(s_{t+1}|s_t,a_t)$ is exactly known for all $(s_t,a_t)\in\mathcal{S}\times\mathcal{A}$; and (b) sufficient computational resources are available to calculate $Q^\pi(s_t,a_t)$  for all $(s_t,a_t)\in\mathcal{S}\times\mathcal{A}$. 
These assumptions are a major impediment for solving process control problems, wherein, complex process behaviour might not be accurately known, or might change over time, and the state and action spaces may be continuous. For an RL solution to be practical, one typically needs to settle for approximate solutions. In the next section, we introduce Q-learning that approximates the optimal Q-function (and thus the optimal policy) using the agent's experiences (or samples) as opposed to process knowledge. Such class of  approximate solutions to the RL problem is called the model-free RL methods. 

\section{Q-learning}
\label{sec:QLearning}
Q-learning is one of the most important breakthroughs in RL  \cite{watkins1989learning, watkins1992q}. 
The idea is to learn  $Q^\star$ directly, instead of first learning $Q^\pi$ 
and then computing $Q^\star$ in (\ref{eq:OAF}). 
Q-learning constructs $Q^\star$ through successive approximations.
Similar to (\ref{eq:BE_AF}), using the Bellman equation, 
$Q^\star$ satisfies the identity
\begin{align}
\label{eq:BE_OAF}
&Q^{\star}(s_t, a_t) = \mathbb{E}_{s_{t+1}\sim p(\cdot|s_t,a_t)}[r(s_t, a_t) + \gamma \max_{a'\in\mathcal{A}}Q^{\star}(s_{t+1}, a')].
\end{align}
There are several ways to use~(\ref{eq:BE_OAF}) to compute $Q^\star$. 
The standard Q-iteration (QI) is a model-based method that requires complete knowledge of the states, the transition function, and the reward function to evaluate the expectation in (\ref{eq:BE_OAF}). 
Alternatively, temporal difference (TD) learning is a model-free method that uses sampling experiences, 
$(s_t,a_t,r_t,s_{t+1})$, to approximate $Q^\star$ \cite{sutton1998reinforcement}. 
This is done as follows.
First, the agent explores the environment by following some stochastic behaviour policy, 
$\beta:\mathcal{S}\rightarrow\mathcal{P}(\mathcal{A})$, 
and receiving an experience tuple $(s_t,a_t,r_t,s_{t+1})$ at each time step. 
The generated tuple is then used to improve the current approximation of $Q^\star$,
denoted $\widehat Q_i$, as follows: 
\begin{align}
\label{eq:TD_Approx_Action_Function}
\widehat{Q}_{i+1}(s_t,a_t)\leftarrow \widehat{Q}_{i}(s_t,a_t) +\alpha\delta,
\end{align}
where $\alpha\in(0,1]$ is the learning rate, 
and $\delta$ is the TD error, defined by
\begin{align}
\delta = r(s_t,a_t)+\gamma\max_{a'\in\mathcal{A}}\widehat{Q}_i(s_{t+1},a') - \widehat{Q}_i(s_t,a_t).
\end{align}
The conditional expectation in (\ref{eq:BE_OAF}) falls away in TD learning in (\ref{eq:TD_Approx_Action_Function}),
since $s_{t+1}$ in $(s_t,a_t,r_t,s_{t+1})$ is distributed according to the target density, 
$p(\,\cdot\,|s_t,a_t)$.
Making the greedy policy choice using $\widehat Q_{i+1}$ instead of $Q^\star$ 
 produces
\begin{align}
\label{eq:Approx_GreedyPolicy}
\widehat{\pi}_{i+1}(s_t)=\argmax_{a\in\mathcal{A}}\widehat{Q}_{i+1}(s_t,a),
\end{align}
where $\widehat{\pi}_{i+1}$ is a greedy policy based on $\widehat{Q}_{i+1}$. 
In the RL literature,  (\ref{eq:TD_Approx_Action_Function}) is referred to as {policy evaluation} 
and (\ref{eq:Approx_GreedyPolicy}) as  {policy improvement}. 
Together, steps (\ref{eq:TD_Approx_Action_Function}) and (\ref{eq:Approx_GreedyPolicy}) 
are called Q-learning \cite{watkins1992q}. 
Pseudo-code for implementing this approach is given in Algorithm \ref{alg:QLearning}. 
Algorithm \ref{alg:QLearning} is a model-free, on-line and {off-policy} algorithm. 
It is model-free as it does not require an explicit model of the environment%
, and on-line because it only utilizes the latest experience tuple 
to implement the policy evaluation and improvement steps.  
Further, Algorithm \ref{alg:QLearning} is off-policy because the agent  
acts in the environment according to its behaviour policy $\beta$, 
but still learns its own policy, $\pi$. Observe that the behaviour policy in 
Algorithm~\ref{alg:QLearning}  is $\epsilon$-greedy, in that it generates greedy actions 
for the most part but has a non-zero probability, $\epsilon$, of generating a random action. 
Note that off-policy is a critical component in RL 
as it ensures a combination of {exploration} and {exploitation}. 
Finally, for Algorithm \ref{alg:QLearning}, 
it can be shown that $\widehat{Q}_i\rightarrow Q^\star$ with probability $1$ as $i\to\infty$ 
\cite{watkins1992q}.

\linespread{1}
\begin{algorithm}[t]
 \algsetup{linenosize=\tiny}
  \caption{\small Q-learning}
  \begin{algorithmic}[1]
  \STATE \textbf{Output:} Action-value function $Q(s,a)$
  \STATE Initialize: Arbitrarily set $Q$, e.g., to $0$ for all states, set $Q$ for terminal states as $0$
\FOR {{each episode}}
\STATE Initialize state $s$ 
\FOR{{each step of episode, state $s$ is not terminal}}
\STATE $a\leftarrow$ action for $s$ derived by $Q$, e.g., $\epsilon$-greedy
\STATE take action $a$, observe $r$ and $s'$
\STATE $\delta \leftarrow r+\gamma\max_{a'} Q(s',a') -Q(s,a)$
\STATE $Q(s,a)\leftarrow Q(s,a)+\alpha\delta$
\STATE $s\leftarrow s'$
\ENDFOR
\ENDFOR  
     \end{algorithmic}
    \label{alg:QLearning}
\end{algorithm}      

\subsection{Q-learning with Function Approximation}
\label{sec:QLearningWithFA}
While Algorithm \ref{alg:QLearning} enjoys strong theoretical convergence properties,
it requires storing the $\widehat{Q}$-values for all state-action pairs $(s,a)\in\mathcal{S}\times\mathcal{A}$.  
In control, where both $\mathcal{S}$ and $\mathcal{A}$ are infinite sets, 
some further simplification is required. 
Several authors have proposed space discretization methods \cite{sutton1996generalization}. 
While such discretization methods may work for simple problems, 
in general, they are not efficient in capturing the complex dynamics of industrial processes.

The problem of {generalizing} Q-learning from finite to continuous spaces 
has been studied extensively  over the last two decades. 
The basic idea in continuous spaces is to use a function approximator, or FA.
A function approximator $Q(s,a,w)$ is a parametric function $Q(s,a,w)$,
whose parameters $w$ are chosen to make $Q(s,a,w) \approx  Q(s,a)$ 
for all $(s\times a)\in\mathcal{S}\times\mathcal{A}$. 
This is achieved by minimizing the following quadratic loss function  $\mathcal{L}$:
\begin{align}
\label{eq:LossFun}
\mathcal{L}_t(w)= \mathbb{E}_{s_t\sim \rho^\beta(\cdot), a_t\sim 
\beta(\cdot|s_t)}\big[(\tilde{y}_t- {Q}(s_t,a_t,w))^2\big],
\end{align}
where $\tilde{y}_t$ is the {target} and $\rho^\beta$ is a discounted state visitation distribution 
under behaviour policy $\beta$. 
The role of $\rho^\beta$ is to weight $\mathcal{L}$ based on how frequently 
a particular state is expected to be visited. 
Further, as in any supervised learning problem, 
the target, $\tilde{y}_t$ is given by $Q^\star(s_t,a_t)$; 
however, since $Q^\star$ is unknown, it can be replaced with its approximation. 
A popular choice of an approximate target is a bootstrap target (or a TD target), given as 
\begin{align}
\label{eq:Target}
\tilde{y}_t = \mathbb{E}_{s_{t+1}\sim 
p(\cdot|s_t,a_t)}[r(s_t,a_t)+\gamma\max_{a'\in\mathcal{A}}{Q}(s_{t+1},a', w)].
\end{align}
In contrast to supervised learning, 
where the target is typically independent of model parameters, 
the target in (\ref{eq:Target}) depends on the FA parameters. 
Finally, (\ref{eq:LossFun}) can be minimized  using a {stochastic gradient descent} (SGD) algorithm. 
An SGD is an iterative optimization method that adjusts $w$ in the direction 
that would most reduce $\mathcal{L}(w)$ for it. 
The update step for SGD is given as follows
\begin{align}
\label{eq:SGD_UpdateRule}
w_{t+1}\leftarrow w_t-\frac{1}{2} \alpha_{c,t}\nabla\mathcal{L}_t(w_t),
\end{align}
where $w_{t}$ and $w_{t+1}$ are the old and new parameter values, respectively, 
and $\alpha_{c,t}$ is a positive step-size parameter. 
Given (\ref{eq:LossFun}), this gradient can be calculated as follows
\begin{align}
\label{eq:LossFunGrad}
\nabla\mathcal{L}_t(w_t)=& -2\mathbb{E}_{s_t\sim \rho^\beta(\cdot), a_t\sim \beta(\cdot|s_t)}\big[\tilde{y}_t- 
{Q}(s_t,a_t,w_t)]\nonumber\\
&\times\nabla_w{Q}(s_t,a_t,w_t).
\end{align}
To derive (\ref{eq:LossFunGrad}), 
it is assumed that $\tilde{y}_t$ is independent of $w$. 
This is a common assumption with TD targets \cite{sutton1998reinforcement}. 
Finally, after updating $w_{t+1}$ (and computing ${Q}(s_{t+1},a',w_{t+1})$), 
the optimal policy can be computed as follows
\begin{align}
\label{eq:Approx_GreedyPolicyFA}
{\pi}(s_{t+1})=\argmax_{a'\in\mathcal{A}}{Q}(s_{t+1},a',w_{t+1}).
\end{align}
The code for Q-learning with FA is given in Algorithm ~\ref{alg:QLearningwithFA}.

\linespread{1}
\begin{algorithm}[t]
 \algsetup{linenosize=\tiny}
  \caption{\small Q-learning with FA}
  \begin{algorithmic}[1]
  \STATE \textbf{Output:} Action value function $Q(s,a,w)$
  \STATE Initialize: Arbitrarily set action-value function weights $w$ (e.g., 
$w = 0$)
\FOR {{each episode}}
\STATE Initialize state $s$ 
\FOR{{each step of episode, state $s$ is not terminal}}
\STATE $a\leftarrow$ action for $s$ derived by $Q$, e.g., $\epsilon$-greedy
\STATE take action $a$, observe $r$ and $s'$
\STATE $\tilde{y} \leftarrow r+\gamma\max_{a'} Q(s',a', w)$
\STATE $w\leftarrow w+ \alpha_{c}(y-Q(s,a,w))\nabla_wQ(s,a,w)$
\STATE $s\leftarrow s'$
\ENDFOR
\ENDFOR  
     \end{algorithmic}
    \label{alg:QLearningwithFA}
\end{algorithm}  

The effectiveness of Algorithm~\ref{alg:QLearningwithFA}  depends on the choice of FA.
Over the past decade, various FAs, both parametric and non-parametric, have been proposed,
including linear basis, Gaussian  processes, radial basis, and Fourier basis. 
For most of these choices, Q-learning with TD targets may be biased \cite{lazaric2010finite}. 
Further, unlike Algorithm \ref{alg:QLearning}, 
the asymptotic convergence of $Q(s,a,w)\rightarrow Q(s,a)$ with Algorithm~\ref{alg:QLearningwithFA} 
is not guaranteed. 
This is primarily due to Algorithm~\ref{alg:QLearningwithFA} using an off-policy 
(i.e., behaviour distribution), bootstrapping (i.e., TD target) and FA approach---a 
combination known in the  RL community as the `{deadly triad}' \cite{sutton1998reinforcement}. 
The possibility of divergence in the presence of the {deadly triad} are well known. 
Several examples have been published: 
see \citet{tsitsiklis1997analysis, baird1995residual, fairbank2011divergence}. 
The root cause for the instability remains unclear---taken one by one, 
the factors listed above are not problematic. 
There are still many open problems in off-policy learning. 
Despite the lack of theoretical convergence guarantees, 
all three elements of the {deadly triad} are also necessary 
for learning to be effective in practical  applications. 
For example, FA is required for scalability and generalization, 
bootstrapping for computational and data efficiency, 
and off-policy learning for decoupling the behaviour policy from the target policy. 
Despite the limitations of Algorithm~\ref{alg:QLearningwithFA}, 
recently, \citet{mnih2015human, mnih2013playing} have successfully adapted Q-learning 
with FAs to learn to play \texttt{Atari} games from pixels. 
For details, see \citet{kober2013reinforcement}.



\section{Policy Gradient}
\label{sec:DPG}
While Algorithm \ref{alg:QLearningwithFA}  generalizes Q-learning to continuous spaces, 
the method lacks convergence guarantees except for with linear FAs, 
where it has been shown not to diverge. 
Moreover, target calculations in (\ref{eq:Target}) and greedy action calculations 
in (\ref{eq:Approx_GreedyPolicyFA}) require maximization of the $Q$-function 
over the action space. 
Such optimization steps are computationally  impractical for large and unconstrained FAs, 
and for continuous action spaces. 
Control applications typically have both these complicating characteristics, 
making Algorithm \ref{alg:QLearningwithFA} is nontrivial to implement.

\linespread{1}
\begin{algorithm}[t]
 \algsetup{linenosize=\tiny}
  \caption{\small Policy Gradient -- REINFORCE}
  \begin{algorithmic}[1]
  \STATE \textbf{Output:} Optimal policy $\pi(a|s,\theta)$
  \STATE Initialize: Arbitrarily set policy parameters $\theta$
\FOR {{true}}
\STATE generate an episode $s_0,a_0,r_1,\dots,s_{T-1},a_{T-1},r_{T}$, following 
$\pi(\cdot|\cdot,\theta)$
\FOR{{each step $t$ of episode $0,1,\dots T-1$}}
\STATE $R_t\leftarrow$ return from step $t$
\STATE $\theta\leftarrow \theta+ 
\alpha_{a,t}\gamma^tR_t\nabla_\theta\pi(a_t|s_t,\theta)$
\ENDFOR
\ENDFOR  
     \end{algorithmic}
    \label{alg:REINFORCE}
\end{algorithm}    

Instead of approximating $Q^\star$ and then computing $\pi$ 
(see Algorithms~\ref{alg:QLearning}   and~\ref{alg:QLearningwithFA}), 
an alternative approach is to directly compute the optimal policy, 
$\pi^\star$ without consulting the optimal $Q$-function. 
The $Q$-function may still be used to learn the policy, 
but is not required for action selection. 
Policy gradient methods are reinforcement Learning algorithms that 
work directly in the policy space. 
Two separate formulations are available:
the average reward formulation and the start state formulation. 
In the start state formulation, the goal of an agent is to obtain a policy $\pi_\theta$, 
parameterized by $\theta\in\mathbb{R}^{n_\theta}$, 
that maximizes the value of starting at state $s_0\in\mathcal{S}$ and following policy $\pi_\theta$. 
For a policy, $\pi_\theta$, the agent's performance  can be evaluated as
\begin{align}
\label{eq:PG_Obj}
J(\pi_\theta)
&= V^{\pi_\theta}(s_0) 
= \mathbb{E}_{h\sim p^\pi(\cdot)}[R_1(h)|s_0],
\end{align}
Observe that the agent's performance in (\ref{eq:PG_Obj}) is completely described 
by the policy parameters in $\theta$. 
A policy gradient algorithm maximizes (\ref{eq:PG_Obj}) by computing an optimal value of $\theta$. 
A stochastic gradient ascent (SGA) algorithm  adjusts $\theta$ 
in the direction of $\nabla_\theta J(\pi_\theta)$, such that 
\begin{align}
\label{eq:PolicyGradient_Iteration}
\theta_{t+1}
\leftarrow \theta_t+\alpha_{a,t}\nabla_\theta J(\pi_\theta)|_{\theta=\theta_t},
\end{align}
where $\alpha_{a,t}$ is the learning rate. 
Calculating  $\nabla_\theta J(\pi_\theta)$ 
requires distribution of states under the current policy, 
which as noted earlier, is unknown for most processes.  
The implementation of policy gradient is made effective by the {policy gradient theorem} 
that calculates a closed-form solution for $\nabla_\theta J(\pi_\theta)$ without reference 
to the state distribution \cite{sutton2000policy}.
The policy gradient theorem establishes that 
\begin{align}
\label{eq:StochasticPolicyGradient1}
&\nabla_\theta J(\pi_\theta) 
= \mathbb{E}_{s_t\sim \rho^\pi_\gamma(\cdot),~ a_t
\sim  \pi_\theta(\cdot|s_t)}\big[Q^{\pi}(s_t,a_t)\nabla_\theta\log{\pi_\theta(a_t|s_t)
}\big],
\end{align}
where $\rho^\pi_\gamma(s) :=\sum_{t=0}^\infty\gamma^{t}p(s_t=s|s_0,\pi_\theta)$ 
is the discounted state visitation distribution and 
$p(s_t|s_0,\pi)$ is a $t$-step ahead state transition density from $s_0$ to $s_t$. 
Equation  (\ref{eq:StochasticPolicyGradient1}) gives a closed-form solution 
for the gradient in (\ref{eq:PolicyGradient_Iteration}) in terms of the $Q$-function 
and the gradient of the policy being evaluated. 
Further, (\ref{eq:StochasticPolicyGradient1}) assumes a stochastic policy 
(observe the expectation over the action space). 
Now, since the expectation is over the policy being evaluated, i.e.. $\pi_\theta$, 
(\ref{eq:StochasticPolicyGradient1}) is an {on-policy} gradient.
Note that an {off-policy} gradient theorem can also be derived 
for a class of stochastic policies. 
See \citet{degris2012off} for details.

To implement (\ref{eq:PolicyGradient_Iteration}) with the policy gradient theorem 
in (\ref{eq:StochasticPolicyGradient1}), 
%
we 
replace the expectation in (\ref{eq:PolicyGradient_Iteration}) with its sample-based estimate, 
and replace $Q^{\pi}$ with the actual returns, $R_t$.  
%
This leads to a policy gradient algorithm, called REINFORCE \cite{williams1992simple}. 
Pseudo-code for REINFORCE is given in Algorithm \ref{alg:REINFORCE}. 
In contrast to Algorithms \ref{alg:QLearning} and \ref{alg:QLearningwithFA}, 
Algorithm \ref{alg:REINFORCE} avoids solving complex optimization problems 
over continuous action spaces and generalizes effectively in continuous spaces. 
Further, unlike Q-learning that always learns a deterministic greedy policy, 
Algorithm \ref{alg:REINFORCE} supports both deterministic and stochastic policies. 
Finally, Algorithm \ref{alg:REINFORCE} exhibits good convergence properties, 
with the  estimate in (\ref{eq:PolicyGradient_Iteration}) guaranteed to converge 
to a local optimum if the estimation of $\nabla_\theta J(\pi_\theta)$ 
is unbiased (see \citet{sutton2000policy}).

Despite the advantages of Algorithm \ref{alg:REINFORCE} over traditional Q-learning,
 Algorithm \ref{alg:REINFORCE} is not amenable to online implementation 
as it requires access to $R_t$ -- the total reward the agent is expected to receive at the end of an episode.  
Further, replacing the $Q$-function by $R_t$ leads to large variance 
in the estimation of $\nabla_\theta J(\pi_\theta)$ \cite{sutton2000policy,  riedmiller2007evaluation}, 
which in turn leads to slower convergence  \cite{konda2000actor}.  
%
An approach to address these issues is to use a low-variance, 
bootstrapped estimate of the $Q$-function, as in the actor-critic architecture.

\begin{figure}[t]
    \centering
    \includegraphics[scale=0.55]{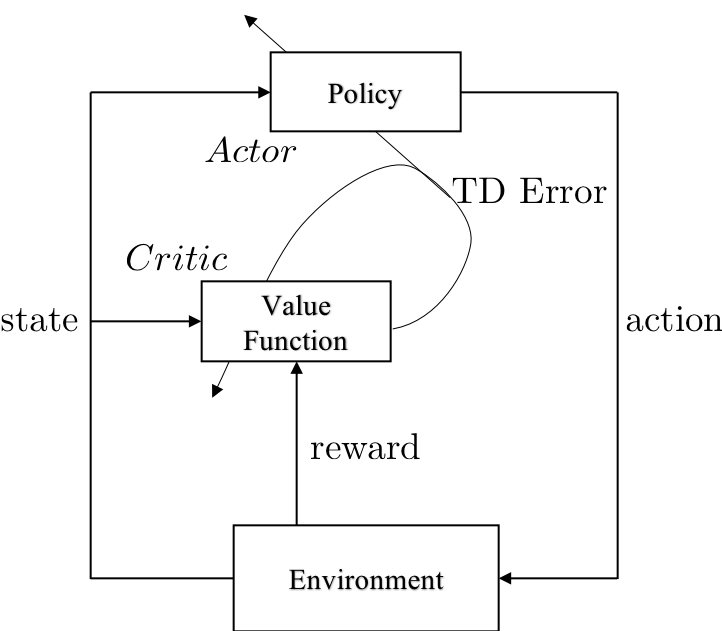}
    \caption{A schematic of the actor-critic architecture}
    \label{fig:ACMethodArchitecture}
\end{figure}

\section{Actor-Critic Architecture}
\label{sec:ActorCritic}
The actor-critic  is a widely used architecture that combines 
the advantages of policy gradient with $Q$-learning 
\cite{sutton2000policy, degris2012off, bhatnagar2008incremental}.
Like policy gradient, actor-critic methods generalize to continuous spaces, 
while the issue of large variance is countered by bootstrapping, 
such as  $Q$-learning with TD update. 
A schematic of the actor-critic architecture is shown in Figure~\ref{fig:ACMethodArchitecture}. 
The actor-critic architecture consists of two eponymous components: an {actor} that  finds an optimal policy, 
and a {critic} that evaluates the current policy prescribed by the actor. 
The actor implements the policy gradient method by adjusting policy parameters using SGA, 
as shown in (\ref{eq:PolicyGradient_Iteration}). 
The critic approximates the $Q$-function in (\ref{eq:StochasticPolicyGradient1}) using an FA. 
With a critic, (\ref{eq:StochasticPolicyGradient1}) can be approximately written as follows: 
\begin{align}
\label{eq:StochasticPolicyGradient_Approx}
&\widehat\nabla_\theta J(\pi_\theta) = \nonumber\\
&\mathbb{E}_{s_t\sim \rho^\pi_\gamma 
(\cdot), a_t\sim \pi_\theta(\cdot|s_t)}\big[Q^\pi(s_t,a_t, 
w)\nabla_\theta\log{\pi_\theta(a_t|s_t)}\big],
\end{align}
where $w\in\mathbb{R}^{n_w}$ is recursively estimated by the critic using SGD in (\ref{eq:SGD_UpdateRule})
and $\theta\in\mathbb{R}^{ n_\theta}$ is recursively estimated by the actor using SGA 
by substituting  (\ref{eq:StochasticPolicyGradient_Approx}) into  (\ref{eq:PolicyGradient_Iteration}). 
Observe that while the actor-critic method combines the policy gradient with $Q$-learning, 
the policy is not directly inferred from $Q$-learning, as in (\ref{eq:Approx_GreedyPolicyFA}). 
Instead, the policy is updated in the policy gradient direction in (\ref{eq:PolicyGradient_Iteration}). 
This avoids the costly optimization in $Q$-learning, 
and also ensures that changes in the $Q$-function only result in small changes in the policy, 
leading  to less or no oscillatory behaviour in the policy. 
Finally, under certain conditions, 
implementing  (\ref{eq:PolicyGradient_Iteration}) with  (\ref{eq:StochasticPolicyGradient_Approx}) 
guarantees that $\theta$ converges  to the local optimal policy \cite{sutton2000policy, degris2012off}.

\subsection{Deterministic Actor-Critic Method}
For a stochastic policy $\pi_\theta$, 
calculating the gradient in (\ref{eq:StochasticPolicyGradient1}) requires integration 
over the space of states and actions. 
As a result, computing the policy gradient for a stochastic policy may require many samples, 
especially in high-dimensional action spaces. 
To allow for efficient calculation of the policy gradient in~(\ref{eq:StochasticPolicyGradient1}), \citet{silver2014deterministic} propose a deterministic policy gradient (DPG) framework.
This assumes that the agent follows a deterministic policy 
$\mu_\theta\colon\mathcal{S}\rightarrow\mathcal{A}$. 
For a deterministic policy, $a_t=\mu_\theta(s_t)$ with probability $1$, 
where  $\theta\in\mathbb{R}^{n_\theta}$ is the policy parameter.
The corresonding performance in (\ref{eq:PG_Obj}) can be written as follows
\begin{align}
\label{eq:DPG_Obj}
J(\mu_\theta)& = V^{\mu_\theta}(s_0)=\mathbb{E}_{h\sim 
p^{\mu}(\cdot)}\bigg[\sum_{t=1}^{\infty} \gamma^{k-1}r(s_t,\mu_\theta(s_t))|s_0 
\bigg].
\end{align}
Similar to the policy gradient theorem in (\ref{eq:StochasticPolicyGradient1}), 
\citet{silver2014deterministic} proposed a {deterministic gradient theorem} 
to calculate the gradient of (\ref{eq:DPG_Obj}) with respect to the  parameter $\theta$: 
\begin{align}
&\nabla_\theta J(\mu_\theta) 
= \mathbb{E}_{s_t\sim \rho^\mu_\gamma(\cdot)}\big[\nabla_a 
Q^\mu(s_t,a)|_{a=\mu_\theta(s_t)}\nabla_\theta\mu_\theta(s_t)\big],
\label{eq:DPGb}
\end{align}
where $\rho^\mu_\gamma$ is the discounted state distribution under policy $\mu_\theta$ 
(similar to $\rho^\pi_\gamma$ in (\ref{eq:StochasticPolicyGradient1})). 
Note that unlike (\ref{eq:StochasticPolicyGradient1}), 
the gradient in  (\ref{eq:DPGb}) only involves expectation over the  states generated according to $\mu_\theta$.
This makes the DPG framework computationally more efficient to implement compared to the stochastic policy gradient.

\linespread{1}
\begin{algorithm}[t]
 \algsetup{linenosize=\tiny}
  \caption{\small Deterministic Off-policy Actor-Critic Method}
  \begin{algorithmic}[1]
  \STATE \textbf{Output:} Optimal policy $\mu_\theta(s)$
  \STATE Initialize: Arbitrarily set policy parameters $\theta$ and $Q$-function 
weights $w$
\FOR {{true}}
\STATE initialize $s$, the first state of the episode
\FOR{$s$ {is not terminal}}
\STATE $a\sim\beta(\cdot|s)$
\STATE take action $a$, observe $s'$ and $r$
\STATE $\tilde{y}\leftarrow r+\gamma Q^\mu(s',\mu_\theta(s'),w)$ 
\STATE $w\leftarrow w+ \alpha_{w}(\tilde{y}-Q^\mu(s,\mu_\theta(s),w))\nabla_w 
Q^\mu(s,a,w)$
\STATE $\theta\leftarrow 
\theta+\alpha_{\theta}\nabla_\theta\mu_\theta(s)\nabla_a 
Q^\mu(s,a,w)|_{a=\mu_\theta(s)}$
\STATE $s\leftarrow s'$
\ENDFOR
\ENDFOR  
     \end{algorithmic}
    \label{alg:ACAlgorithm}
\end{algorithm} 

\begin{figure*}[h]
    \centering
    \includegraphics[scale = 0.4]{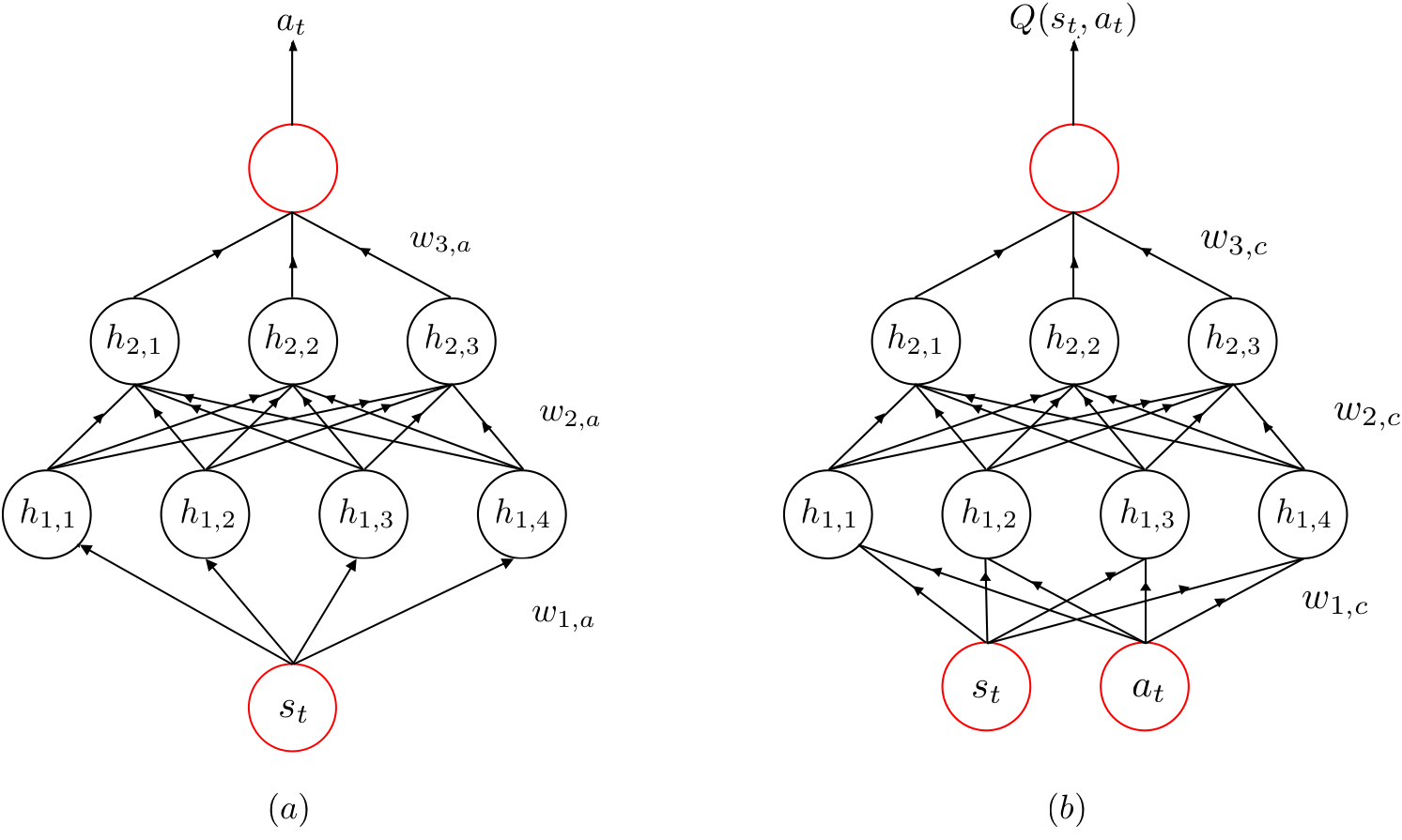}
    \caption{{A deep neural network representation of (a) the actor, and (b) 
the critic.
The red circles represent the input and output layers and the black circles 
represent the hidden layers of the network.}}
    \label{fig:actor_critic}
\end{figure*}

To ensure that the DPG continues to explore the state and action spaces satisfactorily, 
it is possible to implement DPG off-policy.
For a stochastic behaviour policy $\beta$ 
and a deterministic policy  $\mu_\theta$, \citet{silver2014deterministic} showed that a DPG
exists and can be analytically calculated as
\begin{align}
\label{eq:DPGOffPolicy}
&\nabla_\theta J(\mu_\theta) = \mathbb{E}_{s_t\sim 
\rho^\beta_\gamma(\cdot)}\big[\nabla_a 
Q^\mu(s_t,a)|_{a=\mu_\theta(s_t)}\nabla_\theta\mu_\theta(s_t)\big],
\end{align}
where $\rho^\beta_\gamma$ is the discounted state distribution under behaviour 
policy $\beta$.
Compared to (\ref{eq:DPGb}), the off-policy DPG in (\ref{eq:DPGOffPolicy}) 
involves expectation with respect to the states generated by a behaviour policy $\beta$.
Finally, the off-policy DPG can be implemented using the actor-critic architecture.
The policy parameters, $\theta$, can be recursively updated by the actor using 
(\ref{eq:PolicyGradient_Iteration}), where $\nabla_\theta J(\mu_\theta)$ is given in (\ref{eq:DPGOffPolicy}); 
and the $Q$-function, $Q^\mu(s_t,a)$ in (\ref{eq:DPGOffPolicy}) is replaced with a critic, 
$Q^\mu(s_t,a,w)$, whose parameter vector $w$ is recursively estimated using (\ref{eq:SGD_UpdateRule}).
Pseudo-code for the off-policy deterministic actor-critic is given in 
Algorithm~\ref{alg:ACAlgorithm}.

\section{Deep Reinforcement Learning (DRL) Controller}
\label{sec:RLProcessControl}
In this section, we connect the terminologies and methods for RL discussed in  Section \ref{sec:ActorCritic}, and propose a new controller, referred to as a DRL controller. The DRL controller is a model-free controller based on DPG; and is implemented using the actor-critic architecture and uses two independent deep neural networks to generalize the actor and critic to continuous state and action spaces.
\subsection{States, Actions and Rewards}
\label{sec:SAR}

We consider discrete dynamical systems with input and output sequences $u_t$ and $y_t$, respectively.
For simplicity, we focus on the case where the outputs $y_t$ contain full information on the state of
the system to be controlled.
Removing this hypothesis is an important practical element of our ongoing research in this area.

The correspondence between the agent's action in the RL formulation
and the plant input from the control perspective is direct: we identify $a_t=u_t$.
The relationship between the RL state and the state of the plant is subtler.
The RL state, $s_t\in\mathcal{S}$, must capture all the features of the environment 
on the basis of which the RL agent acts. 
To ensure that the agent has access to relevant process information, 
we define the RL state as a tuple of the current and past outputs, 
past actions and current deviation from the set-point $y_{\rm sp}$, such that
\begin{align}
\label{eq:RLState}
s_t := \langle y_t, \dots, y_{t-d_y}, a_{t-1},\dots,a_{t-d_a},(y_t-y_{\rm sp})\rangle,
\end{align}
where $d_y\in\mathbb{N}$ and $d_a\in\mathbb{N}$ denote the number of past output and input values, respectively. \textcolor{black}{In this paper, it is assumed that $d_y$ and $d_a$ are known a priori.}   Note that for $d_a=0$, $d_y=0$, the RL state is `memoryless', in that
\begin{align}
\label{eq:RLStateMemoryless}
s_t := \langle y_t ,(y_t-y_{\rm sp})\rangle.
\end{align}
Further, 
we explore only deterministic policies expressed using a single-valued function
$\mu\colon\mathcal{S}\rightarrow\mathcal{A}$, 
so that for each state $s_t\in\mathcal{S}$, 
the probability measure on $\mathcal{A}$ defining the next action puts weight $1$
on the single point $\mu(s_t)$.

\color{black}

The goal for the agent in a set-point tracking problem  is to find an optimal policy, 
$\mu$, that reduces the tracking error. 
This objective is incorporated in the RL agent by means of a reward function 
$r\colon\mathcal{S}\times\mathcal{A}\times\mathcal{S}\rightarrow\mathbb{R}$,
whose aggregate value the agent tries to maximize. 
In contrast with MPC, where the controller minimizes the tracking error over the space of control actions, 
here the agent maximizes the reward it receives over the space of policies. 
We consider two reward functions.

The first -- an {$\ell_1$-reward function} -- measures the negative $\ell_1$-norm of the tracking error. 
Mathematically, for a multi-input and multi-output (MIMO) system with $n_y$
outputs, the $\ell_1$-reward is
\begin{equation}
\label{eq:rewardFn}
r(s_t,a_t,s_{t+1}) = -\sum_{i=1}^{n_y} |y_{i,t}-y_{i,{\rm sp}}|,
\end{equation}
where $y_{i,t}\in\mathcal{Y}$ are the $i$-th output, $y_{i,{\rm sp}}\in\mathcal{Y}$
is the set-point for the $i$-th output. 
Variants of the $\ell_1$-reward function are presented and discussed in 
Section~\ref{sec:Implementation}. 
\color{black} 
The second reward -- a {polar reward function} -- assigns a $0$ reward if the
tracking error is a monotonically decreasing function at each sampling time for
all $n_y$ outputs or $-1$ otherwise.
Mathematically, the polar reward function is
\begin{align}
\label{eq:polar_reward}
&r(s_t,a_t,s_{t+1}) =\nonumber\\
&\begin{cases}
0 \quad \text{if } |y_{i,t}-y_{i,sp}| > |y_{i,t+1}-y_{i,sp}|\quad \forall
i\in\{1,\dots, n_y\}\\
-1    \quad \text{otherwise}
\end{cases}
\end{align}
Observe that a polar reward (\ref{eq:polar_reward}) incentivizes gradual improvements in
tracking performance, which leads to less aggressive control strategy and a
smoother tracking compared to the $\ell_1$-reward in (\ref{eq:rewardFn}).

\color{black}
\subsection{Policy and Q-function Approximations}
We use neural networks to generalize the policy and the $Q$-function to continuous state and action spaces (see \ref{sec:Basics_NN} for basics of neural networks). The  policy, $\mu$, is represented using a deep feed-forward neural network,  parameterized by weights $W_a\in\mathbb{R}^{n_a}$, such that given
$s_t\in\mathcal{S}$ and $W_a$, the policy network produces an output  $a_t=\mu(s_t, W_a)$. Similarly, the $Q$-function is also represented using a deep neural network,
parameterized by weights $W_c\in\mathbb{R}^{n_c}$, such that given $(s_t,a_t)\in\mathcal{S}\times\mathcal{A}$ and $W_c$, the $Q$-network outputs $Q^\mu(s_t,a_t, W_c)$.

\begin{figure*}[h]
	\centering
	\includegraphics[scale=0.3]{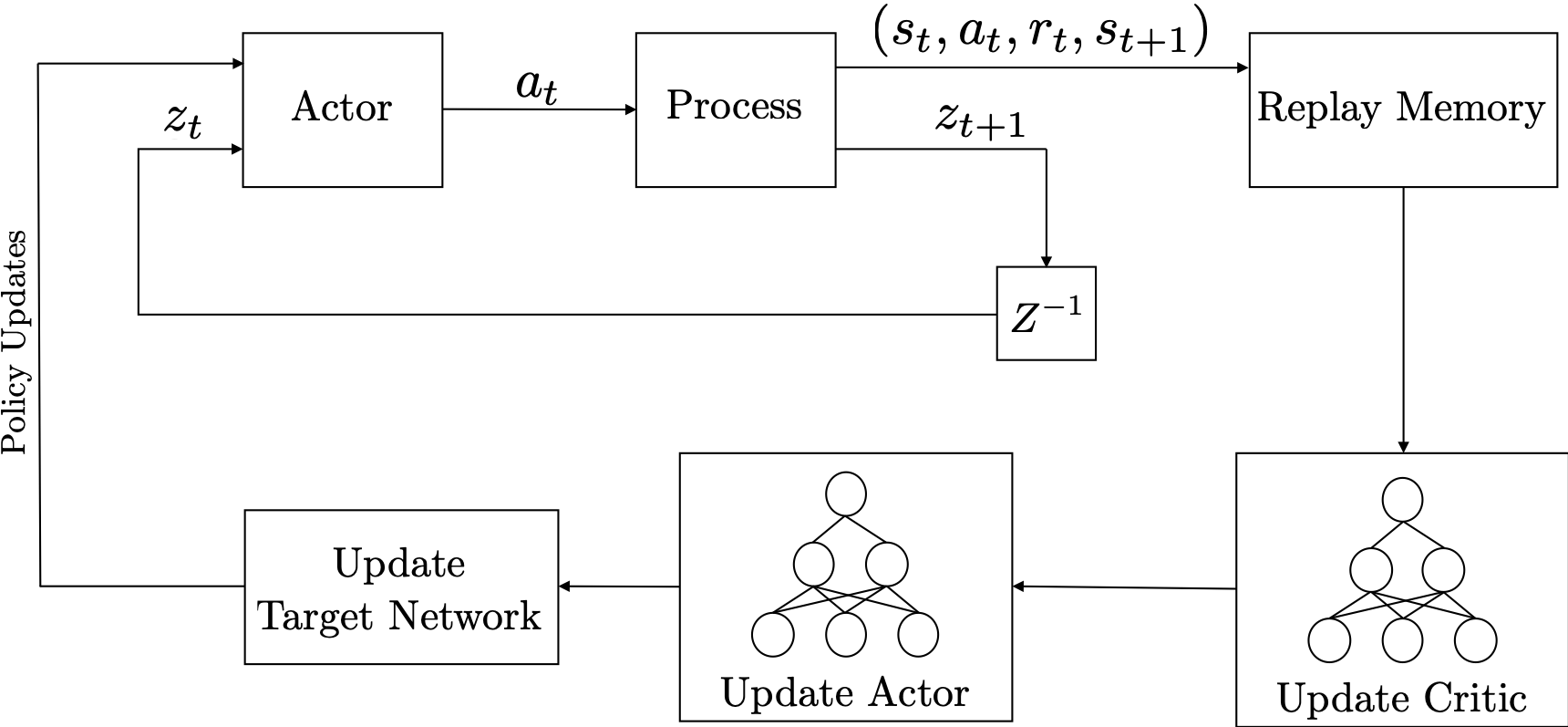}
	\caption{{A schematic of the proposed DRL controller architecture for set-point tracking problems.}}
	\label{fig:learning_overview}
\end{figure*}

\subsection{Learning Control Policies}
As noted, the DRL controller uses a deterministic off-policy actor-critic architecture to learn the control policy in set-point tracking problems.
The proposed method is similar to Algorithm \ref{alg:ACAlgorithm} or the one proposed by \citet{lillicrap2015continuous}, but modified for set-point tracking  problems. In the proposed DRL controller architecture, the actor is represented by $\mu(s_t, W_a)$ and a critic is represented by $Q(s_t,a_t, W_c)$. A schematic of the proposed DRL controller architecture is illustrated in
Figure \ref{fig:actor_critic}. The critic network predicts $Q$-values for each state-action pair, and the  actor network proposes an action for a given state.
The goal is then to learn the actor and critic neural network parameters by  interacting with the process plant. Once the networks are trained, the  actor network is  used to compute the optimal action for any given state.

For a given actor-critic architecture, the network parameters can be readily estimated using SGD (or SGA); 
however, these methods  are not effective in applications, 
where the state, action and reward sequences  are temporally correlated. 
This is because  SGD provides a $1$-sample network  update, 
assuming that the samples are independently and identically distributed (iid).
For dynamic systems, for which a tuple $(s_t,a_t,r_t,s_{t+1})$ may be temporally correlated to the past tuples 
(e.g.,  $(s_{t-1},a_{t-1},r_{t-1},s_{t})$), up to the Markov order, 
such iid network updates in presence of correlations are not effective.

To make learning more effective for dynamic systems, 
we propose to break the  inherent temporal correlations between the tuples by randomizing it. 
To this effect, we use a batch SGD, rather than a single sample SGD for network training. 
As in DQN, we use a {replay memory} (RM) for batch training of the  networks. 
The RM is a finite memory that stores a a large number of tuples, 
denoted as  $\{(s^{(i)},a^{(i)},r^{(i)},s'^{(i)})\}_{i=1}^K$, where $K$ is the size of the  RM. 
As a queue data structure, the latest tuples are always stored in the RM, 
and the old tuples are discarded to keep the cache size constant.
At each time, the network is updated by uniformly sampling  $M$ ($M\leq K$) tuples  from the RM. 
The critic update in Algorithm \ref{alg:ACAlgorithm} with RM  results in a batch stochastic gradient, 
with the following update step
\begin{align}
\label{eq:QwithRM}
W_c\leftarrow & W_c+
\frac{\alpha_{c}}{M}\sum_{i=1}^M(\tilde{y}^{(i)}-Q(s^{(i)},\mu(s^{(i)},W_a),W_c))\nonumber\\
&\times \nabla_{W_c}Q^\mu(s^{(i)},\mu(s^{(i)},W_a),W_c),
\end{align}
where
\begin{align}
\label{eq:TargetRM0}
\tilde{y}^{(i)} \leftarrow r^{(i)}+\gamma Q^\mu(s'^{(i)},\mu(s'^{(i)},W_a), W_c),
\end{align}
for all $i=1,\dots, M$. 
The batch update, similar to (\ref{eq:QwithRM}), can also be derived for the  actor network using an RM.

We propose another strategy to further stabilize the actor-critic architecture. 
First, observe that the parameters of the critic network, $W_c$, 
being updated  in (\ref{eq:QwithRM})  are also used in calculating the target, $\tilde{y}$ in  (\ref{eq:TargetRM0}). 
Recall that, in supervised learning, the actual target is independent of $W_c$;  
however, as discussed in Section \ref{sec:QLearningWithFA}, 
in absence of  actual targets, we use $W_c$ in (\ref{eq:TargetRM0}) to approximate the target  values. 
Now, if the $W_c$ updates in (\ref{eq:QwithRM}) are erratic, 
then the target  estimates in (\ref{eq:TargetRM0}) are also erratic, 
and may cause the network to  diverge, as observed in \citet{lillicrap2015continuous}. 
We propose to use a separate network, called {target network} to estimate  the target in (\ref{eq:TargetRM0}).  
Our solution is similar to the target network used in \citet{lillicrap2015continuous, mnih2013playing}.
Using a target network, parameterized by $W_c'\in\mathbb{R}^{n_c}$, 
(\ref{eq:TargetRM0}) can be written as follows
 \begin{align}
\label{eq:TargetRM}
\tilde{y}^{(i)} \leftarrow r^{(i)}+\gamma Q^\mu(s'^{(i)},\mu(s'^{(i)},W'_a), W'_c),
\end{align}
where
\begin{align}
\label{eq:TargetNetwork}
W_c'\leftarrow\tau W_c +(1-\tau)W_c',
\end{align}
 and $0<\tau<1$ is the target network update rate. 
Observe that the parameters of the target network in (\ref{eq:TargetNetwork})  
are updated by having them slowly track the parameters of the critic network, $W_c$.  
In fact, (\ref{eq:TargetNetwork}) ensures that the target values change  slowing, 
thereby improving the stability of learning. A similar target network can also be used for stabilizing the actor network.

Note that while we consider the action space to be continuous, in practice, it  is also often bounded. 
This is because the controller actions typically involve changing bounded  physical quantities, 
such as flow rates,  pressure and pH. To ensure that the actor network produces  feasible controller actions, 
it is  important to bound the network over the feasible action space.
Note that if the networks are not constrained then the critic network will continue providing gradients 
that encourage the actor network to take actions  beyond the feasible space.

In this paper, we assume that the action space, $\mathcal{A}$ is an interval action space, 
such that $\mathcal{A}=[a_L,a_H]$, where $a_L<a_H$ (the  inequality holds element-wise for systems with multiple inputs). 
One approach to enforce the constraints on the network is to bound the output  layer of the actor network. 
This is done by {clipping} the gradients used by the actor network in the  update step 
(see Step $10$ in Algorithm \ref{alg:ACAlgorithm}). 
For example, using the clipping gradient method proposed in  \cite{hausknecht2015deep}, 
the gradient used by the actor, $\nabla_a Q^\mu(s,a,w)$, can be clipped as follows
\begin{align}
&\nabla_a Q^\mu(s,a,w)\leftarrow \nabla_a Q^\mu(s,a,w)\nonumber\\
&\times
\begin{cases}
(a_{H}-a)/(a_{H}-a_{L}) ,& \text{if } \nabla_a Q^\mu(s,a,w) \text{~increases~}
a \\
(a-a_{L})/(a_{H}-a_{L}),              & \text{otherwise}
\end{cases} \label{eq:Gradients}
\end{align}
Note that in (\ref{eq:Gradients}), the gradients are down-scaled as the  controller action approaches 
the boundaries of its range, and are inverted if  the controller action exceeds the range. 
With (\ref{eq:Gradients}) in place, even if the critic continually recommends  increasing the controller action, it will converge to the its upper bound,  $a_H$. Similarly, if the critic decides to decrease the controller action, it will  decrease immediately. 
Using (\ref{eq:Gradients}) ensures that the controller actions are within the  feasible space. 
Putting all the improvement strategies discussed in this section, 
a  schematic of the DRL controller architecture for set-point tracking 
is  shown in Figure \ref{fig:learning_overview}, and the code is provided in  Algorithm \ref{alg:ActorCriticProposed}.

\linespread{1}
\begin{algorithm}[h]
 \algsetup{linenosize=\tiny}
  \caption{\small Deep Reinforcement Learning Controller}
  \begin{algorithmic}[1]
  		\STATE \textbf{Output:} Optimal policy $\mu(s,W_a)$
 		\STATE Initialize: $W_a, W_c$ to random initial weights
 		\STATE Initialize: $W'_a\gets W_a$ and $W'_c\gets W_c$
		\STATE Initialize: Replay memory with random policies
		\FOR{{each episode}}{}{}
		
		\STATE Initialize an output history $\langle y_{0}, \ldots, y_{-n+1} \rangle$ from an action history $\langle a_{-1}, \ldots, a_{-n} \rangle$
		\color{black}
		\STATE  Set $y_{\rm sp}\gets$ set-point from the user
		\FOR{{each step $t$ of episode $0,1,\dots T-1$}}{}{}
		
		\STATE Set $s \gets  \langle y_t, \dots, y_{t-d_y}, a_{t-1},\dots,a_{t-d_a},(y_t-y_{\rm sp})\rangle$
		\STATE Set $a_t\gets \mu(s,W_a) + \mathcal{N}$
		\STATE Take action $a_t$, observe $y_{t+1}$ and $r$
		\STATE Set $s' \gets  \langle y_{t+1}, \dots, y_{t+1-d_y}, a_{t},\dots,a_{t+1-d_a},(y_{t+1} - y_{\rm sp})\rangle$
		\color{black}
		\STATE  Store tuple $(s, a_t, s', r)$ in RM
		\STATE  Uniformly sample $M$ tuples from RM
		\FOR{{$i=1$ to $M$}}{}{}
			\STATE  Set $\tilde{y}^{(i)} \leftarrow r^{(i)}+\gamma
Q^\mu(s'^{(i)},\mu(s'^{(i)},W'_a), W'_c)$
		\ENDFOR
		\STATE Set $W_c\leftarrow W_c+
\frac{\alpha_{c}}{M}\sum_{i=1}^M(\tilde{y}^{(i)}-Q^\mu(s^{(i)},a^{(i)},W_c))\nabla_{W_c}Q^\mu(s^{(i)},a^{(i)},W_c)$
		\FOR{{$i=1$ to $M$}}{}{}
				\STATE Calculate $\nabla_{a}
Q^\mu(s^{(i)},a,W_c)|_{a=a^{(i)}}$
			\STATE Clip $\nabla_{a}
Q^\mu(s^{(i)},a,W_c)|_{a=a^{(i)}}$ using  (\ref{eq:Gradients})
		\ENDFOR
		\STATE Set $W_a\leftarrow
W_a+\frac{\alpha_{a}}{M}\sum_{i=1}^M\nabla_{W_a}\mu(s^{(i)},W_a)\nabla_{a} Q^\mu(s^{(i)},a,W_c)|_{a=a^{(i)}}$
		\STATE Set $W'_a \gets \tau W_a + (1-\tau) W'_a$
		\STATE Set $W'_c \gets \tau W_c + (1-\tau) W'_c$
	\ENDFOR
		\ENDFOR
     \end{algorithmic}
    \label{alg:ActorCriticProposed}
\end{algorithm}

As outlined in Algorithm \ref{alg:ActorCriticProposed}, first we randomly initialize the parameters of the actor and critic networks (Step 2). We then create a copy of the network parameters and initialize the parameters  of the target networks (Step 3).  The final initialization step is to supply the RM with a large enough collection of tuples  $(s, a, s', r)$ on which to begin training the RL controller (Step 4). Each episode is preceded by two steps. First, we ensure the state definition $s_t$ in \eqref{eq:RLState} is defined for the first $d_y$ steps by initializing a sequence of actions  $\langle a_{-1}, \ldots, a_{-n} \rangle$ along with the corresponding sequence of outputs $\langle y_{0}, \ldots, y_{-n+1} \rangle$ where $n \in \mathbb{N}$ is sufficiently large. A set-point is then fixed for the ensuing episode (Steps 6-7). For a given user-defined set-point, the actor, $\mu(s,W_a)$ is queried,  and a control action is implemented on the process (Steps 8--10). Implementing $\mu(s,W_a)$ on the process generates a new output $y_{t+1}$. The controller then receives a reward $r$, which is the last piece of information needed to store the updated tuple $(s, a, r, s')$  in the RM (Steps 9--12). $M$ uniformly sampled tuples are generated from the RM  to update the actor and  critic networks using batch gradient methods (Steps 14--23). Finally, for a fixed $\tau$, we update the target actor and target critic  networks in Steps 24--25. The above steps are then repeated until the end of the episode.

To ensure that proposed method adequately explores the state and action spaces,  the behavior policy for the agent is constructed by adding noise sampled from a  noise process, $\mathcal{N}$,  to our actor policy, $\mu$ (see Step $10$ in  Algorithm \ref{alg:ActorCriticProposed}). Generally,  $\mathcal{N}$ is to be chosen to suit the process.
We use a zero-mean Ornstein-Uhlenbeck (OU) process \cite{uhlenbeck1930theory} to  generate temporally correlated exploration samples; however, the user can  define their own noise process. We also allow for random initialization of  system outputs at the start of an episode to ensure that the policy is not  stuck in a local optimum. Finally, note that Algorithm \ref{alg:ActorCriticProposed} is a fully automatic algorithm that learns the control policy in real-time  by continuously interacting with the process.

\subsection{Network Structure and Implementation}
\label{sec:Implementation}

The actor and critic neural networks each had $2$ hidden layers with $400$ and
$300$ units, respectively. We initialized the actor and critic networks using uniform Xavier initialization \cite{XavierInitialization}. Each unit was modeled using a  Rectified non-linearity activation function. Example \ref{sec:Example3} differs slightly: the output layers were initialized from a uniform distribution over $[-0.003, 0.003]$, and we elected to use a $\tanh$ activation for the second hidden layer.
For all examples, the network hidden layers were {batch-normalized} using the method in
\citet{ioffe2015batch} to ensure that the training is effective in processes where variables have different physical units and scales. Finally, we implemented the Adam optimizer in \citet{kingma2014adam} to train the networks and regularized the weights and biases of the second hidden layer with an $L_2$ weight decay of $0.0001$.
\color{black}

Algorithm \ref{alg:ActorCriticProposed} was scripted in \texttt{Python} and
implemented on an \texttt{iMac} (3.2 GHz Intel Core i5, 8 GB RAM).
 The deep networks for the actor and critic were built using
\texttt{Tensorflow} \cite{abadi2016tensorflow}. For Example \ref{sec:Example3}, we trained the networks on \texttt{Amazon
g2.2xlarge EC2 instances}; the matrix multiplications were performed on a Graphics Processing Unit (GPU)
available in the \texttt{Amazon g2.2xlarge EC2 instances}. The hyper-parameters for Algorithm \ref{alg:ActorCriticProposed} are process-specific and need to be selected carefully. We list in Table \ref{tab:Hyperparameters} the nominal values for hyper-parameters used in all of our numerical examples (see Section \ref{sec:Simulation}); additional specifications are listed at the end of each example. An optimal selection of hyper-parameters is crucial, but is beyond the scope of the current work.

\renewcommand{\arraystretch}{1.0}
\begin{table}[t]
\small
\centering
\caption{The hyper-parameters used in Algorithm $5$.}
\label{tab:Hyperparameters}
	\begin{tabular}{|l|c|c|}
		\hline
		{Hyper-parameter}   &{Symbol} & {Nominal value} \\ \hline
		Actor learning rate & $\alpha_a$ & $10^{-4}$\\
		Critic learning rate & $\alpha_c$ & $10^{-4}$\\
		Target network update rate & $\tau$ & $10^{-3}$\\
		OU process parameters & $\mathcal{N}$ & $\theta = .15, \sigma = .30$\\
		\hline
	\end{tabular}
\end{table}

\subsection{DRL Controller Tuning}
In this section, we provide some general guidelines and recommendations for  implementing the DRL controller (see Algorithm  \ref{alg:ActorCriticProposed}) for set-point tracking problems.

\renewcommand{\labelenumi}{(\alph{enumi})}
\begin{enumerate}[wide=0pt]
\item\textbf{Episodes:}  To begin an episode, we randomly sample an action from the the action space and let the system settle under that action before starting the time steps in the main algorithm. \color{black} For the examples considered in Section
\ref{sec:Simulation}, each episode is terminated after $200$ time steps or when $|y_{i,t}-y_{i,sp}| \leq
\varepsilon$ for all $1 \leq i \leq n_y$ for $5$ consecutive time steps, where $\varepsilon$ is a user-defined tolerance. For processes with large time constants, it is recommended to run the episodes
longer to ensure that the transient and steady-state dynamics are
captured in the input-output data.

\item\textbf{Rewards:} We consider the reward hypotheses (\ref{eq:rewardFn}) and (\ref{eq:polar_reward}) in our examples. A possible variant of (\ref{eq:rewardFn}) is given as follows

\begin{align}
\label{eq:rewardFn_epsilon}
r(s_t,a_t,s_{t+1}) =
\begin{cases}
c \quad \text{if } |y_{i,t}-y_{i,sp}|\leq \varepsilon\quad\forall
i\in\{1,2,\dots,n_y\}\\
-\sum_{i=1}^{n_y}|y_{i,t}-y_{i,sp}|              \quad\text{otherwise}.
\end{cases}
\end{align}

where $y_{i,t}\in\mathcal{Y}$ are the $i$-th output, $y_{i,sp}\in\mathcal{Y}$
is the set-point for the $i$-th output, $c\in\mathbb{R}_{+}$ is a constant
reward, and $\epsilon\in\mathbb{R}_+$ is a user-defined tolerance. According to (\ref{eq:rewardFn_epsilon}), the agent receives the maximum reward, $c$, only if all
the outputs are within the tolerance limit set by the user. The potential advantage of \eqref{eq:rewardFn_epsilon} over (\ref{eq:rewardFn}) is that it can lead to faster tracking. On the other hand, the control strategy learned by the RL agent is generally more aggressive. Ultimately, we prefer the $\ell_1$-reward function due the gradual increasing behavior of the function as the RL agent learns. Note that under this reward hypothesis the tolerance $\varepsilon$ in \eqref{eq:rewardFn_epsilon} should be the same as the early termination tolerance described previously.
\color{black}
These hypotheses are well-suited for set-point tracking  as they use tracking error to generate rewards; however, in other problems, it is imperative to explore other reward hypothesis. Some examples, include --  (i) negative of $2$-norm of the control error instead of $1$-norm in (\ref{eq:rewardFn});  (ii) weighted rewards for different outputs in a MIMO system; (iii) economic reward function.

\item\textbf{Gradient-clipping:} The gradient-clipping in (\ref{eq:Gradients})
ensures that the DRL controller outputs  are feasible and within the operating  range. Also, defining tight lower and upper limits on the inputs has a significant effect on the learning rate. In general, setting tight limits on the inputs lead to faster convergence of  the  DRL controller.

\item\textbf{RL state:}
The definition of an RL state is critical as it affects the performance of the DRL controller.
As shown in (\ref{eq:RLState}), we use a tuple of current and past outputs,
past actions and current deviation from the set-point as the RL state.
While (\ref{eq:RLState}) is well-suited for set-point tracking problems,
it is instructive to highlight that the choice of an RL state is not unique,
as there may be other RL states for which the DRL controller could have
a similar or better performance.
In our experience, including additional relevant information in the RL states
improves the overall performance and convergence-rate of the DRL controller.
The optimal selection of an RL state is not within the scope of the paper;
however, it certainty is an important area of research that warrants additional investigation.

\item\textbf{Replay memory:}  We initialize the RM with $M$ tuples $(s, a, s', r)$ generated by simulating the system response under random actions, where $M$ is the batch size used for training the networks. \color{black} During initial learning, having a large
RM is essential as it gives DRL controller access to the data generated
from the old policy. Once the actor and critic networks are trained, and the system attains
steady-state, the RM need not be updated as the input-output data no
longer contributes new information to the RM.

\item \textbf{Learning:} The DRL controller learns the policy in real-time and
continues to refine it as new experiences are collected.
To reduce the computational burden of updating the actor and critic networks at
each sampling time, it is recommended to `{turn-off}' learning once the
set-point tracking is complete.
For example,  if the difference between the output and the
set-point is within certain predefined threshold, the networks need not be
updated.  We simply terminate the episode once the DRL controller has tracked the set-point for 5 consecutive time steps.

\color{black}

\item\textbf{Exploration:} Once the actor and critic networks are trained, it
is recommended that the agent stops exploring the state and action spaces.
This is because adding exploration noise to a trained network adds unnecessary
noise to the control actions as well as the system outputs.

\end{enumerate}

\color{black}

\section{Simulation Results}
\label{sec:Simulation}
The efficacy of the DRL controller outlined in Algorithm 
\ref{alg:ActorCriticProposed} is demonstrated on four simulation examples.
The first three examples, include:  a single-input-single-output (SISO) 
paper-making machine; a multi-input-multi-output (MIMO) high purity 
distillation column; and a nonlinear MIMO heating, ventilation, and air 
conditioning (HVAC) system.
The fourth example evaluates the robustness of Algorithm 
\ref{alg:ActorCriticProposed} to process changes.


\subsection{Example 1: Paper Machine}
\label{sec:Example1}
In the pulp and paper industry, a paper machine is used to form the paper sheet and then remove water from the sheet by various means, such as vacuuming, pressing, or evaporating.
The paper machine is divided into two main sections: wet-end and dry-end.
The sheet is formed in the wet-end on a continuous synthetic fabric, and the 
dry-end of the machine removes the water in the sheet through evaporation.
The paper passes over rotating steel cylinders, which are  heated by
super-heated pressurized steam. Effective control of the moisture content in the sheets is an active area of 
research.

 \begin{figure*}[t]
	\centering
	\includegraphics[width = \textwidth]{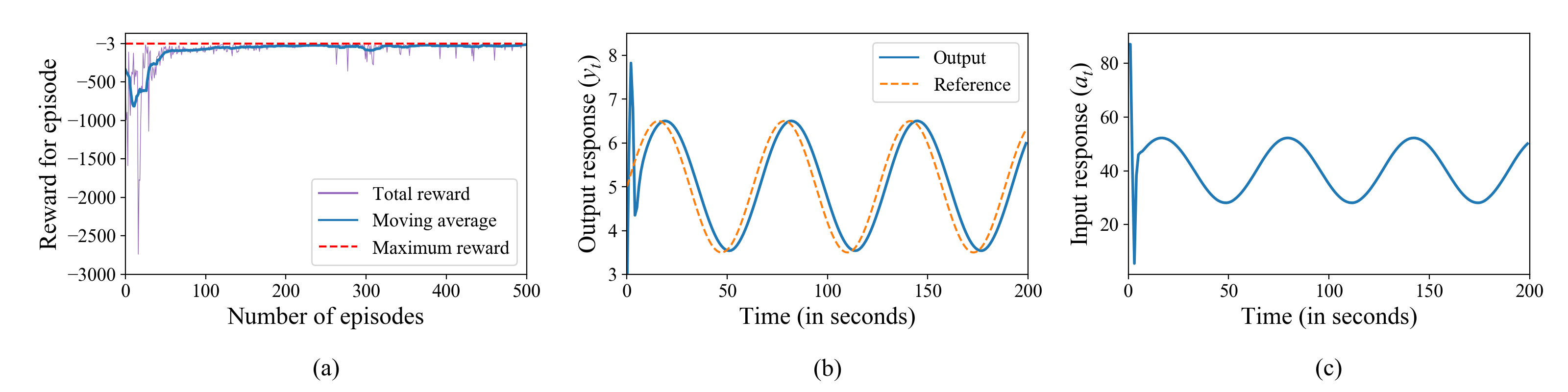}
	\caption{Simulation results for Example $1$ -- (a) moving average with window size 21 (the number of distinct set-points used in training) for the total reward per episode; (b) tracking 
performance of the DRL controller on a sinusoidal reference signal; and (c) the input signal generated by the DRL controller.}
	\label{fig:Example1_fig1}
\end{figure*}

In this section, we design a DRL controller to control the 
moisture content in the sheet, denoted by $y_t$ (in percentage), to a desired 
set-point, $y_{sp}$.
There are several variables that affect the drying time of the sheet, such as 
machine speed, steam pressure, drying fabric, etc.
For simplicity, we only consider the effect of steam flow-rate, denoted by 
$a_t$ (in $m^3/hr$)  on $y_t$.
For simulation purposes, we assume that $a_t$ and $y_t$ are related according 
to the following discrete-time transfer function
\begin{equation}
\label{eq:siso2} 
G(z)=\frac{0.05z^{-1}}{1 - 0.6 z^{-1} }.
\end{equation}
Note that (\ref{eq:siso2}) is strictly used  for generating episodes, and is 
not used  in the design of the DRL controller. 
In fact, the user has complete freedom to substitute (\ref{eq:siso2}) with any linear or nonlinear process model, in case of simulations; or with the actual  data from the paper machine, in case of industrial implementation. Next, we implement Algorithm \ref{alg:ActorCriticProposed} with the 
hyper-parameters listed in Tables \ref{tab:Hyperparameters}--\ref{tab:Hyperparameters1}.

 \begin{figure*}[t]
	\centering
	\includegraphics[width =.85\textwidth]{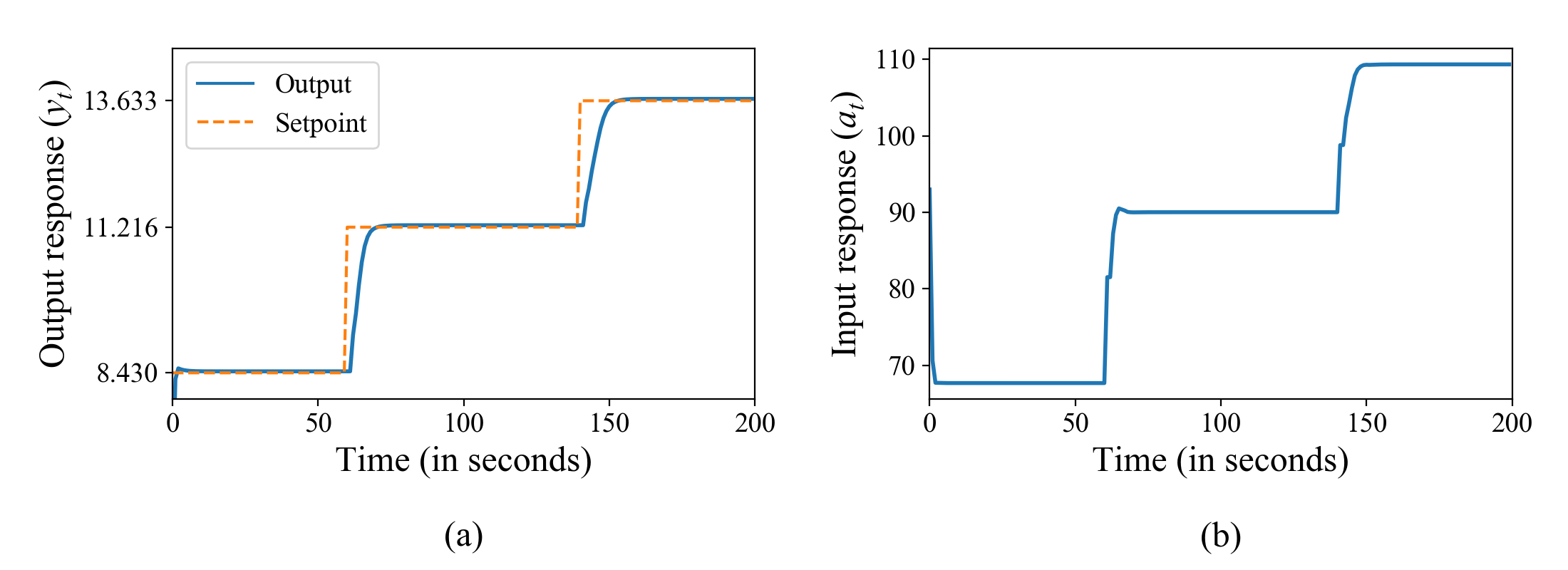}
	\caption{Simulation results for Example $1$ -- (a) tracking performance of the DRL controller on randomly generated set-points outside of the training interval $[0, 10]$; (b) the input signal generated by the DRL controller.}
	\label{fig:Example1_fig2}
\end{figure*}

The generated data is then used for real-time training of the actor and critic 
networks in the DRL controller. Figure \ref{fig:Example1_fig1}(a) captures the increasing trend of the cumulative reward per episode by showing the moving average across the number of set-points used in training.
From Figure \ref{fig:Example1_fig1}(a), it is clear that as the DRL controller 
interacts with the process, it learns a progressively better policy, leading to 
higher rewards per episode. In less than 100 episodes, the DRL controller is able to learn a policy suitable for basic set-point tracking; it was able to generalize effectively in the remaining 400 episodes (see Figures \ref{fig:Example1_fig1}(b)-(c) and \ref{fig:Example1_fig2}).

Figures \ref{fig:Example1_fig1}(b)-(c) showcase the DRL controller, putting aside the physical interpretation of our system \eqref{eq:siso2}. We highlight the smooth tracking performance of the DRL controller on a sinusoidal reference signal in Figure \ref{fig:Example1_fig1}(b), as this exemplifies the robustness of the DRL controller to tracking problems far different from the fixed collection of set-points in the interval $[0, 10]$ seen during training.

Figure \ref{fig:Example1_fig2}(a) shows that the DRL controller is able to track randomly generated set-points with little overshoot, both inside and outside the training output space $\mathcal{Y}$. Finally, noting that our action space for training is $[0, 100]$, Figure \ref{fig:Example1_fig2}(b) shows the DRL controller is operating outside the action space to track the largest set-point.

\renewcommand{\arraystretch}{1.0}
\begin{table}[t]
\small
\centering
\caption{Specifications for Example $1$.}
\label{tab:Hyperparameters1}
	\begin{tabular}{|l|c|c|}
		\hline
		{Hyper-parameter}   &{Symbol} & {Nominal value} \\ \hline
		Episodes & & 500\\
		Mini-batch size & $M$ &128\\
		Replay memory size & $K$ & $5\times10^{4}$\\
		Reward discount factor & $\gamma$ & $0.99$\\
		Action space & $\mathcal{A}$ & $[0, 100]$\\
		Output space & $\mathcal{Y}$ & $[0, 10]$\\
		\hline
	\end{tabular}
\end{table}

\noindent\emph{Further implementation details:}
We define the states as in \eqref{eq:RLStateMemoryless} and select the $\ell_1$-reward in (\ref{eq:rewardFn}). Concretely,
\begin{equation}
r(s_t,a_t,s_{t+1}) = - |y_t - y_{sp}|,
\end{equation}
where $y_{sp}$ is uniformly sampled from $\{0, 0.5, 1, \ldots, 9.5, 10\}$ at the beginning of each episode. We added zero-mean Gaussian noise with variance .01 to the outputs during training. We defined the early-termination tolerance for the episodes to be $\varepsilon = .01$.

\subsection{Example 2: High-purity Distillation Column}
\label{sec:Example2}
\begin{figure}[t]
\centering
\includegraphics[scale=0.4]{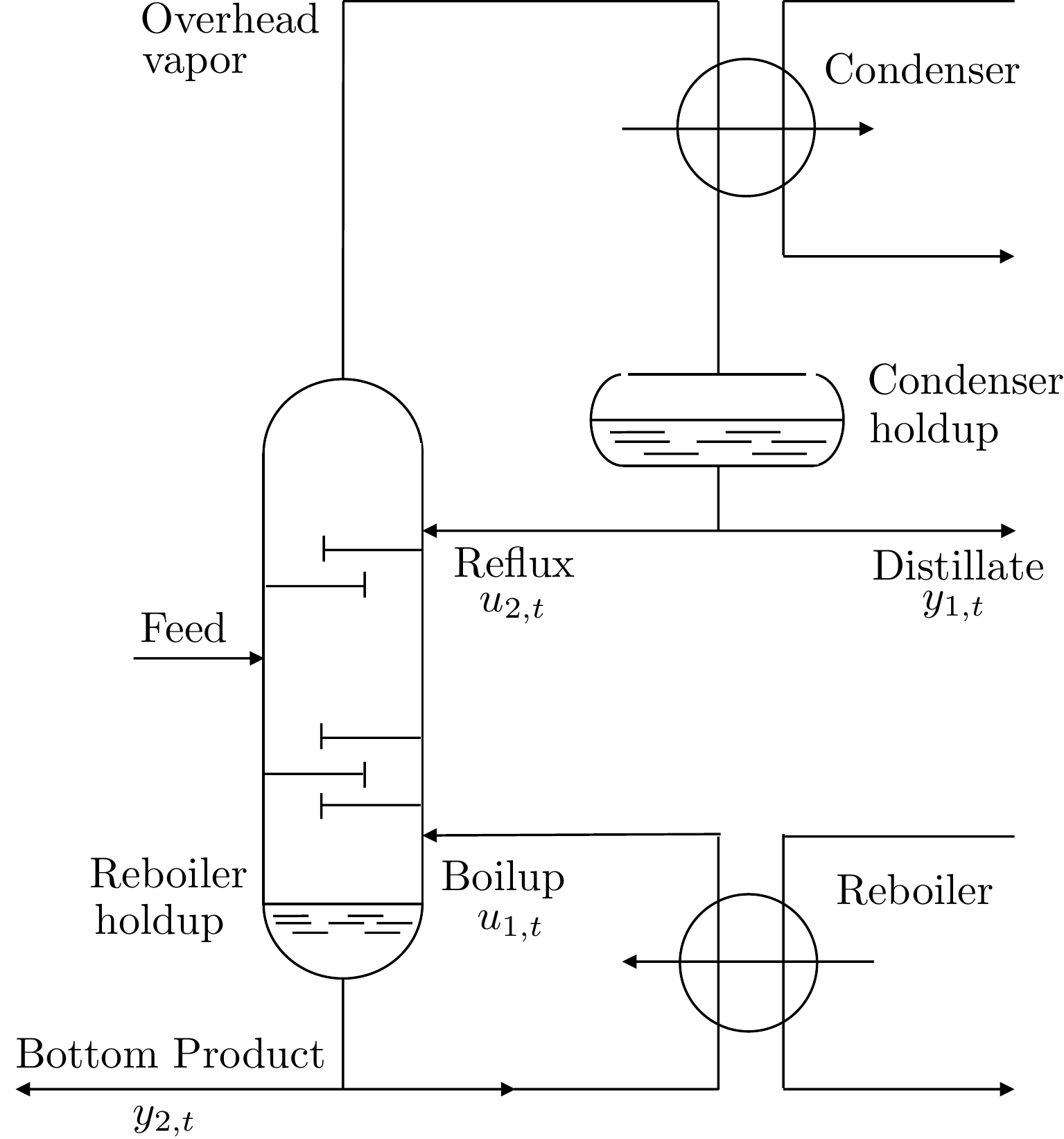}%
\caption{A schematic of a two-product distillation column in Example 
$2$.
Adapted from \cite{skogestad1988robust}.}%
\label{fig:DistillationColumn}
\end{figure}

In this section, we consider a two-product distillation column as shown in 
Figure \ref{fig:DistillationColumn}.
The objective of the distillation column is to split the feed, which is a 
mixture of a light and a heavy component, into a distillate product, $y_{1,t}$, 
and a bottom product, $y_{2,t}$.
 The distillate contains most of the light component and the bottom product 
contain most of the heavy component.
Figure \ref{fig:DistillationColumn} has multiple manipulated variables; 
however, for simplicity, we only consider two inputs: boilup rate, $u_{1,t}$ 
and the reflux rate, $u_{2,t}$.
Note that the boilup and reflux rates have immediate effect on the product 
compositions.
The distillation column shown in Figure \ref{fig:DistillationColumn} has a 
complex dynamics; however, for simplicity, we linearize the model 
around the steady-sate value.
The model for the distillation column is given as follows 
\cite{skogestad1988robust}
\begin{align}
\label{eq:mimo}
G(s) = \begin{bmatrix} 
\displaystyle
\frac{0.878}{\tau s + 1} & \displaystyle -\frac{0.864}{\tau s + 1} \\
&\\
\displaystyle \frac{1.0819}{\tau s + 1} & \displaystyle -\frac{1.0958}{\tau s + 
1} 
\end{bmatrix} \text{, where } \tau > 0.
\end{align}
Further, to include measurement uncertainties, we add a zero-mean Gaussian 
noise to (\ref{eq:mimo}) with variance $0.01$.
The authors in \cite{skogestad1988robust} showed that a distillation column 
with a model representation in (\ref{eq:mimo}) is ill-conditioned.
In other words,  the plant gain strongly depends on the input direction, such 
that inputs in directions corresponding to high plant gains are strongly 
amplified by the plant, while inputs in directions corresponding to low plant 
gains are not.
The issues around controlling ill-conditioned processes are well known to the 
community \cite{waller2005multi,mollov2004analysis}.

\begin{figure*}[h]
\centering
\includegraphics[width =.75\textwidth]{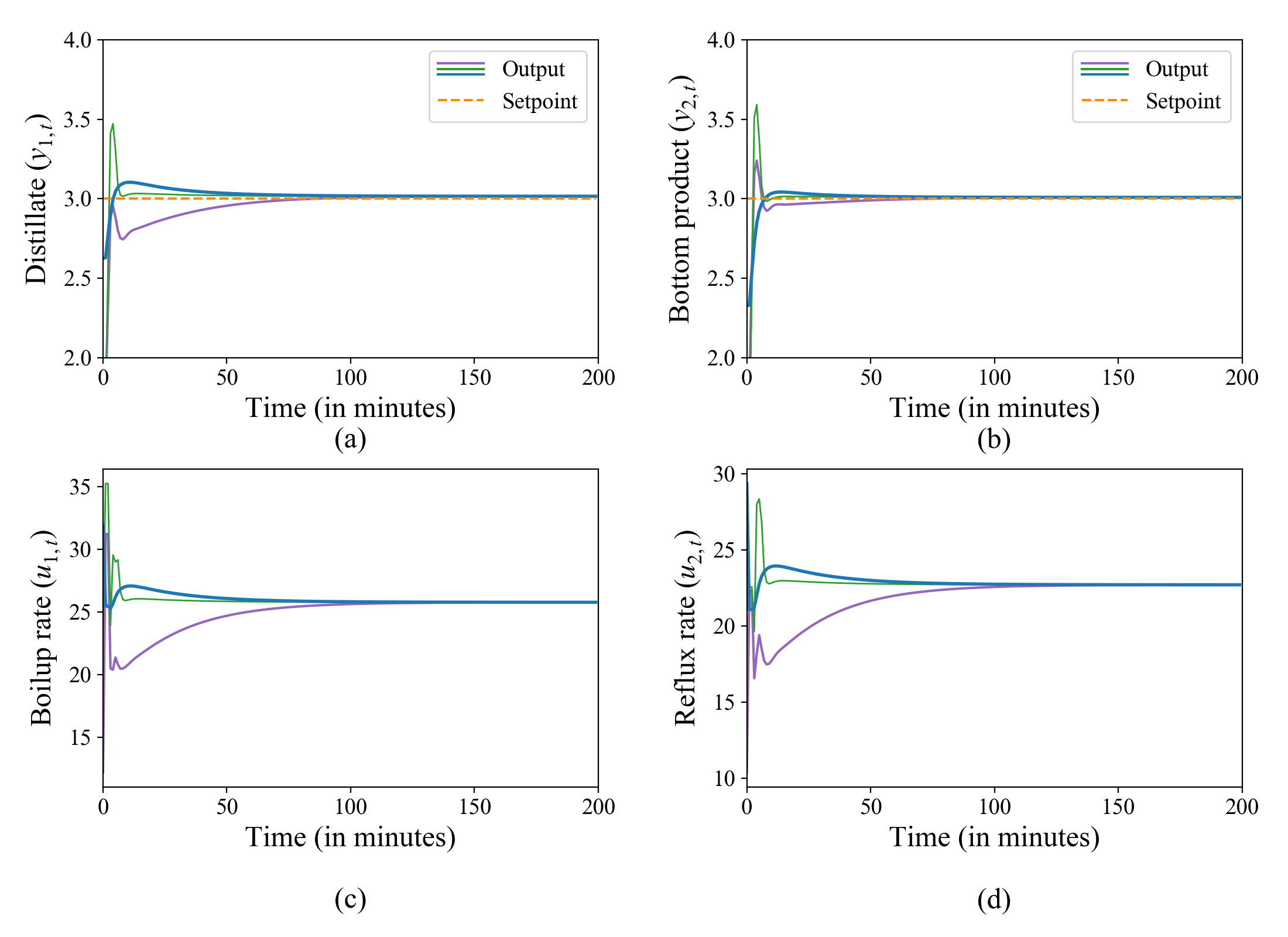}%
\caption{Simulation results for Example $2$ -- (a) and (b) 
set-point tracking for distillate and bottom product, respectively, where the different colors correspond to different starting points; (c) and (d) 
boilup and reflux rates selected by the DRL controller, respectively.}
\label{fig:Example2}
\end{figure*}

Next, we implement Algorithm 
\ref{alg:ActorCriticProposed} with the reward hypothesis in  
(\ref{eq:rewardFn}).
Figures \ref{fig:Example2}(a) and (b) show the performance of the DRL 
controller in tracking the distillate and the bottom product to desired 
set-points for a variety of random initial starting points.
Observe that starting from some initial condition, the DRL controller 
successfully tracks the set-points within $t=75$ minutes.
Figures \ref{fig:Example2}(c) and (d) show the boilup and reflux rates selected 
by the DRL controller, respectively.
This example, clearly highlights the efficacy of the DRL controller in 
successfully tracking a challenging, ill-conditioned MIMO system.
Finally, it is instructive to highlight that the model in (\ref{eq:mimo}) is 
strictly for generating episodes, and is not used by the DRL controller.
The user has freedom to replace (\ref{eq:mimo}) either with a complex 
model, or with the actual plant data.

\begin{figure*}[htbp]
\centering
\includegraphics[width =.75\textwidth]{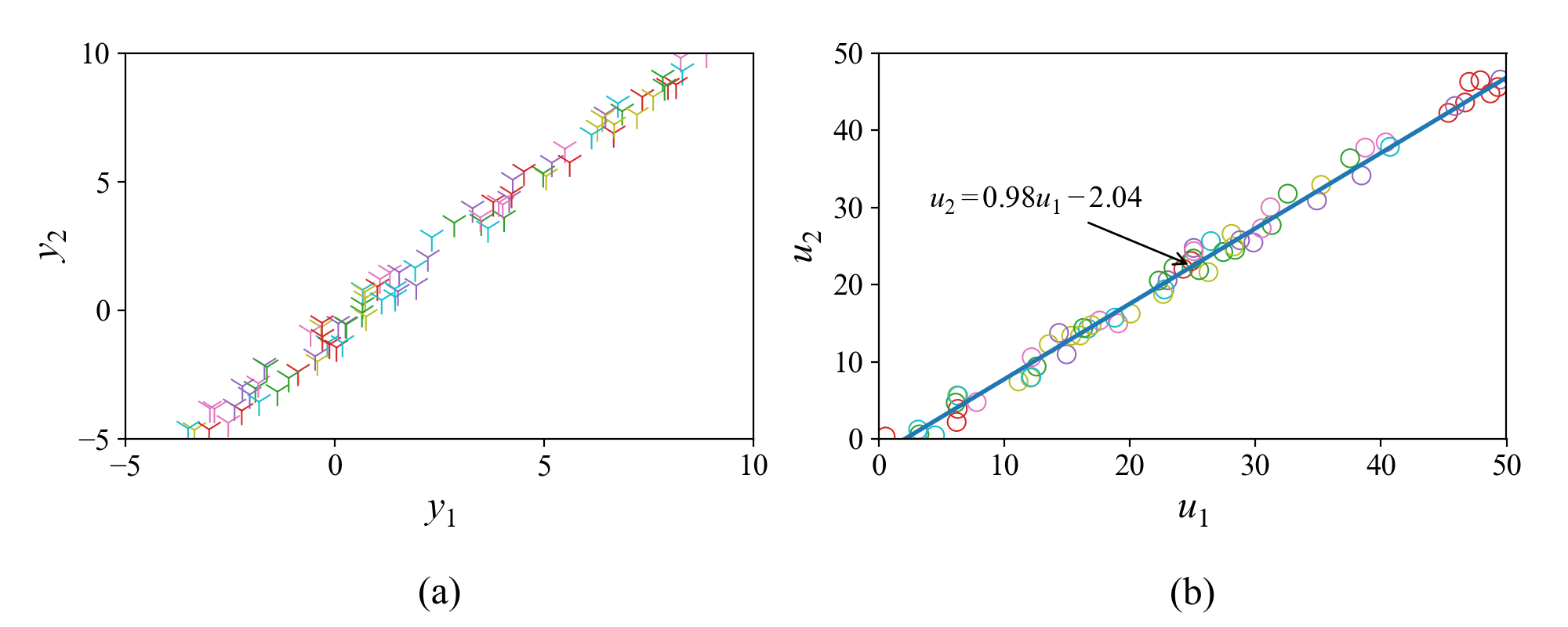}%
\caption{Set-point selection for Example \ref{sec:Example2} -- (a) the distribution of settled outputs after simulating the MIMO system \eqref{eq:mimo} with uniformly sampled actions in $[0, 50] \times [0, 50]$;  (b) the corresponding action pairs such that the settled outputs are both within $[0, 5]$, along with a linear regression of these samples.}
\label{fig:Example2_fig2}
\end{figure*}

\emph{Further implementation details:}
First, note our system \eqref{eq:mimo} needs to be discretized when implementing Algorithm \ref{alg:ActorCriticProposed}.
We used the $\ell_1$-reward hypothesis given in \eqref{eq:rewardFn} and defined the RL state as in (\ref{eq:RLStateMemoryless}).
To speed up learning, it is important to determine which pairs of set-points make sense in practice. To address this, we refer to Figure \ref{fig:Example2_fig2}. We uniformly sample constant input signals from $[0, 50] \times [0, 50]$ and let the MIMO system \eqref{eq:mimo} settle for 50 time steps. Figure \ref{fig:Example2_fig2}(a) shows these settled outputs. Clearly, we have very little flexibility when initializing feasible set-points in Algorithm \ref{alg:ActorCriticProposed}. Therefore, we only select set-points pairs $(y_{1, sp}, y_{2, sp})$ where $y_{1, sp}, y_{2, sp} \in \{0, .5, 1, \ldots, 4.5, 5\}$ and $| y_{1, sp} - y_{2, sp}| \leq .5$ Finally, Figure \ref{fig:Example2_fig2}(a) shows the outputs in a restricted subset of the plane, $[-5, 10] \times [-5, 10]$; the outputs can far exceed this even when sampling action pairs within the action space. A quick way to initialize an episode such that the outputs begin around the desired output space, $[0, 5] \times [0, 5]$, is to initialize action pairs according to a linear regression such as the one shown in Figure \ref{fig:Example2_fig2}(b); we also added zero-mean Gaussian noise with variance 1 to this line. We reiterate that these steps are implemented purely to speed up learning, as the DRL controller will saturate and accumulate a large amount of negative reward when tasked with set-points outside the scope of Figure \ref{fig:Example2_fig2}(a). Further, this step does not solely rely on the process model \eqref{eq:mimo}, as similar physical insights could be inferred from real plant data or by studying the reward signals associated with each possible set-point pair as Algorithm \ref{alg:ActorCriticProposed} progresses. Figure \ref{fig:Example4_fig2} at the end of Example \ref{sec:Example3} demonstrates a possible approach to determining feasible set-points and effective hyper-parameters.
 
\renewcommand{\arraystretch}{1.0}
\begin{table}[t]
\small
\centering
\caption{Specifications for Example $2$.}
\label{tab:Hyperparameters2}
	\begin{tabular}{|l|c|c|}
		\hline
		{Hyper-parameter}   &{Symbol} & {Nominal value} \\ \hline
		Episodes & & 5,000\\
		Mini-batch size & $M$ &128\\
		Replay memory size & $K$ & $5\times10^{5}$\\
		Reward discount factor & $\gamma$ & $0.95$\\
		Action space & $\mathcal{A}$ & $[0, 50]$\\
		Output space & $\mathcal{Y}$ & $[0, 5]$\\
		\hline
	\end{tabular}
\end{table}

\subsection{Example 3: HVAC System}
\label{sec:Example3}
Despite the advances in research on HVAC control algorithms, most field 
equipment is controlled using classical methods, such as hysteresis/on/off and 
Proportional Integral and Derivative (PID) controllers.
Despite their popularity, these classical methods do not perform optimally. The high thermal inertia of buildings induces large time delays in the building 
dynamics, which cannot be handled efficiently by the simple on/off controllers.
Furthermore, due to the high non-linearity in building dynamics coupled with 
uncertainties such as weather, energy pricing, etc., these PID controllers 
require extensive re-tuning or auto-tuning capabilities \cite{wang2001pid}, 
which increases the difficulty and complexity of the control problem 
\cite{wang2017long}.

\begin{figure*}[!ht]
\centering
\includegraphics[width =.75\textwidth]{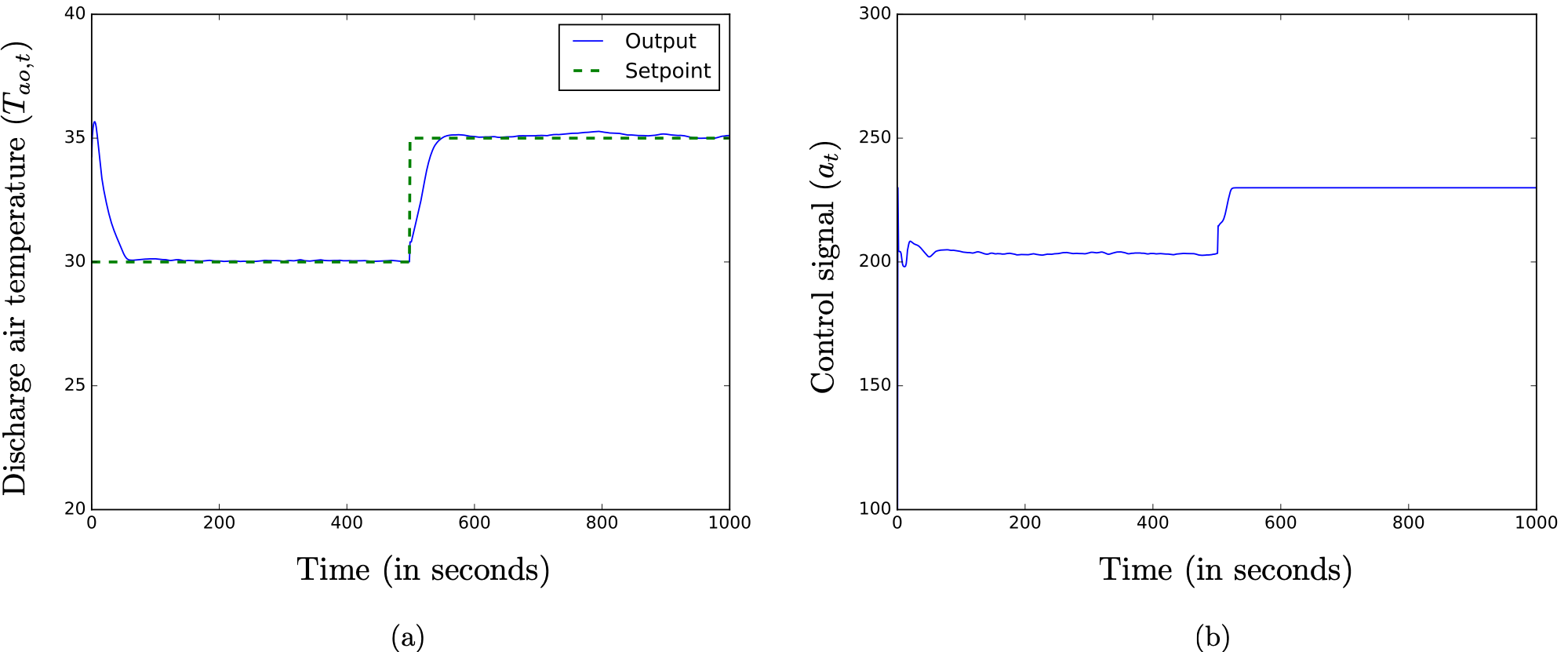}%
\caption{Simulation results for Example $3$ -- (a) the response 
of the discharge air temperature to the set-point change and (b) the control 
signal output calculated by the DRL controller.}
\label{fig:Example3}
\end{figure*}

Due to these challenging aspects of HVAC control, various advanced control 
methods have been investigated, ranging from non-linear model predictive 
control \cite{moradi2011nonlinear} to optimal control 
\cite{greensfelder2011investigation}.
Between these methods, the quality of control performance relies heavily on 
accurate process identification and modelling; however, large variations exist 
with building
design, zone layout, long-term dynamics and wide ranging operating conditions.
In addition, large disturbance effects from external weather, occupancy 
schedule changes and varying energy prices make process identification a very 
challenging problem \cite{wang2017long}.

In this section we demonstrate the efficacy of the proposed \emph{model-free} 
DRL controller to control the HVAC system.
First, to generate episodes, we assume that the HVAC system can be modelled as 
follows \cite{underwood1991dynamic}:
\begin{subequations}
\begin{align}
&T_{wo,t} =  T_{wo,t-1}+\theta_1f_{w,t-1}(T_{wi,t-1}-T_{wo,t-1})\nonumber\\
&+\big(\theta_2+\theta_3f_{w,t-1}+\theta_4f_{a,t-1}\big)[T_{ai,t-1}-\bar{T}_{w,t-1}],\label{eq:HVACModel1}\\
&T_{ao,t} =  T_{ao,t-1}+\theta_5f_{a,t-1}(T_{ai,t-1}-T_{ao,t-1})\nonumber\\
&+(\theta_6+\theta_7f_{w,t-1}+\theta_8f_{a,t-1})(\bar{T}_{w,t-1}-T_{ai,t-1})\nonumber\\
&+\theta_9(T_{ai,t}-T_{ai,t-1}),\label{eq:HVACModel2}\\
&\bar{T}_{w,t}=0.5[T_{wi,t}+T_{wo,t}]\label{eq:HVACModel3},
\end{align}
\end{subequations}
where $T_{wo}\in\mathbb{R}_+$ and $T_{ao}\in\mathbb{R}_+$ are the outlet water 
and discharge air temperatures (in $^\circ$ C), respectively; 
$T_{wi}\in\mathbb{R}_+$ and $T_{ai}\in\mathbb{R}_+$ are the inlet water and air 
temperatures (in $^\circ$ C), respectively; $f_w\in\mathbb{R}_+$ and 
$f_a\in\mathbb{R}_+$ are the water and air mass flow rates (in kg/s), 
respectively; and $\theta_i\in\mathbb{R}$ for $i=1,\dots, 9$ are various 
physical parameters, with nominal values given in \cite{underwood1991dynamic}.
The water flow rate, $f_w$, is assumed to be related to the controller action 
as follows:
\begin{align}
\label{eq:ValveModel}
f_{w,t} = \theta_{10}+\theta_{11}a_{t}+\theta_{11}a^2_{t}+\theta_{12}a^3_{t},
\end{align}
where $a_t$ is the control signal in terms of $12$-bits and $\theta_j$, for 
$j = 10, 11, 12$ are valve model parameters given in \cite{underwood1991dynamic}.

In (\ref{eq:HVACModel1})--(\ref{eq:HVACModel3}), $T_{wi}$, $T_{ai}$ and $f_a$ 
are disturbance variables that account for the set-point changes and other 
external disturbances. Also, $T_{wi}$, $T_{ai}$ and $f_a$ are assumed to follow constrained 
random-walk models, such that
\begin{align}
\label{eq:Constraints}
0.6\leq f_{a,t}\leq 0.9; \quad 73\leq T_{wi,t}\leq 81; \quad 4\leq T_{ai,t}\leq 
10,
\end{align}
for all $t\in\mathbb{N}$.
For simplicity, we assume that the controlled variable is $T_{ao}$ and the 
manipulated variable is $f_w$.
The objective is to design a DRL controller to achieve desired discharge air 
temperature, $y_{t}\equiv T_{ao,t}$ by manipulating $a_t$.

Next, we implement Algorithm \ref{alg:ActorCriticProposed} to train the DRL 
controller for the HVAC system.
Using the $\ell_1$-reward hypothesis in (\ref{eq:rewardFn}) we observed a 
persistent offset in $T_{ao}$ to set-point changes.
Despite increasing the episode length, the DRL controller was still unable to 
eliminate the offset with (\ref{eq:rewardFn}).
Next, we implement Algorithm \ref{alg:ActorCriticProposed} with the polar 
reward hypothesis in (\ref{eq:polar_reward}).
With (\ref{eq:polar_reward}), the response to control actions was found to be 
less noisy and \emph{offset-free}.
In fact, Algorithm \ref{alg:ActorCriticProposed} performed much better with (\ref{eq:polar_reward}) in terms of noise and offset reduction, compared 
to (\ref{eq:rewardFn}).
Figure \ref{fig:Example3}(a) shows the the response of the discharge air 
temperature to a time-varying set-point change.
Observe that the DRL controller is able to track the set-point changes in less 
than $50$ seconds with \emph{small} offset.
Moreover, it is clear that the  response to the change in control action with 
(\ref{eq:polar_reward}) is smooth.
Finally, the control signal generated by the DRL controller is shown in Figure 
\ref{fig:Example3}(b).
This example demonstrates the efficacy of Algorithm 
\ref{alg:ActorCriticProposed} in learning the  control policy of complex 
nonlinear processes simply by interacting with the process in real-time.

\renewcommand{\arraystretch}{1.0}
\begin{table}[h]
\small
\centering
\caption{Specifications for Example $3$.}
\label{tab:Hyperparameters3}
	\begin{tabular}{|l|c|c|}
		\hline
		{Hyper-parameter}   &{Symbol} & {Nominal value} \\ \hline
		Episodes & & 100,000\\
		Mini-batch size & $M$ &64\\
		Replay memory size & $K$ & $10^6$\\
		Reward discount factor & $\gamma$ & $0.99$\\
		Action space & $\mathcal{A}$ & $[150, 800]$\\
		Output space & $\mathcal{Y}$ & $[30, 35]$\\
		\hline
	\end{tabular}
\end{table}

\begin{figure*}[!ht]
\centering
\includegraphics[width =.75\textwidth]{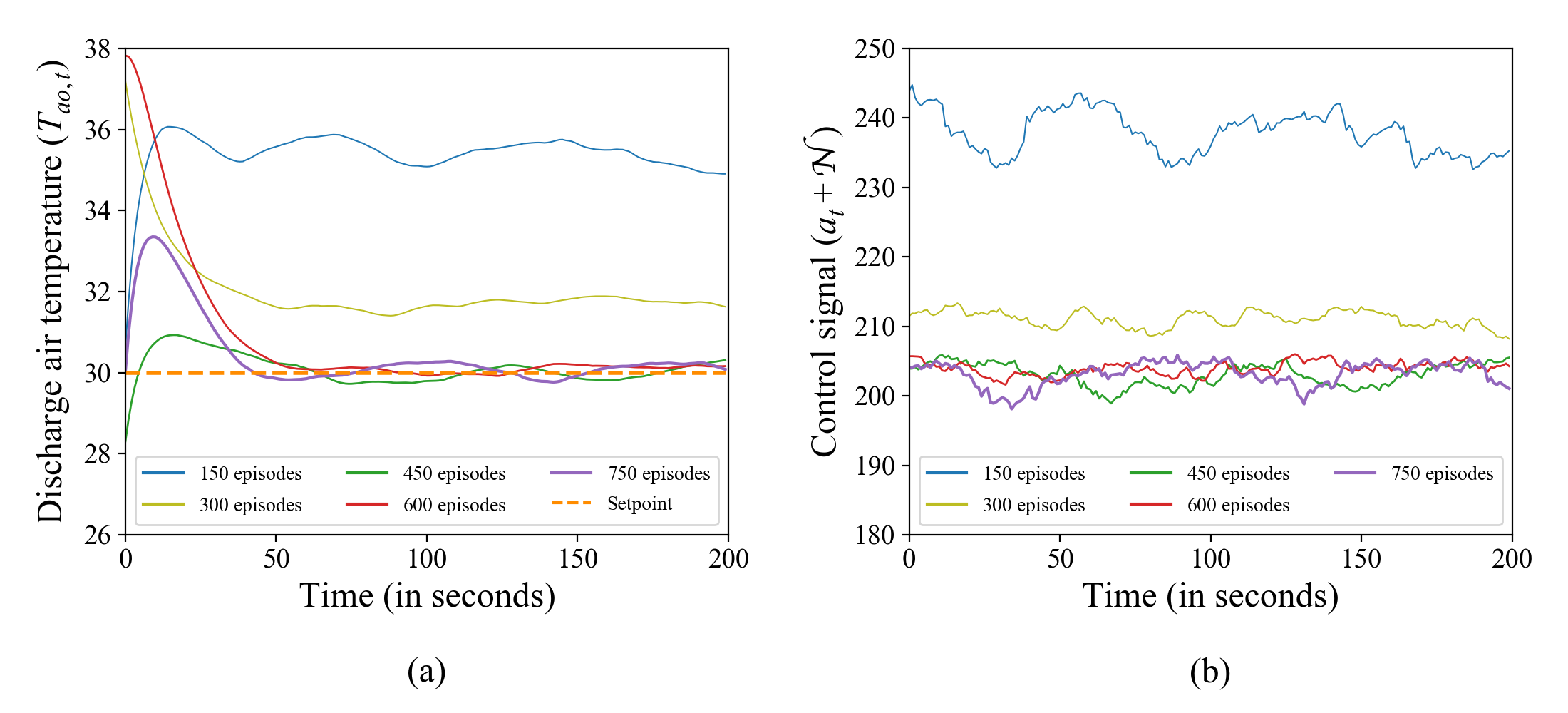}%
\caption{Snapshots during training of the input and output signals -- (a) progression of set-point 
tracking performance of the DRL controller; (b) the respective controller actions taken by the DRL controller.}
\label{fig:Example4_fig2}
\end{figure*}

\emph{Further implementation details:} We contrast the  long training time listed in Table \ref{tab:Hyperparameters3} to generate Figure \ref{fig:Example3} by noting from Figure \ref{fig:Example4_fig2} that the DRL controller was able to learn to track a fixed set-point, $y_{sp} = 30$, in fewer than 1000 episodes. In general, training the DRL controller with a single set-point is a reasonable starting point for narrowing the parameter search with Algorithm \ref{alg:ActorCriticProposed}.

\subsection{Example 4: Robustness to Process Changes}
\label{sec:Example4}
In this section, we evaluate the robustness of DRL controller to adapt to 
abrupt process changes.
 We revisit the example from the pulp and paper industry in Section 
\ref{sec:Example1}.

\begin{figure*}[!ht]
\centering
\includegraphics[width =.75\textwidth]{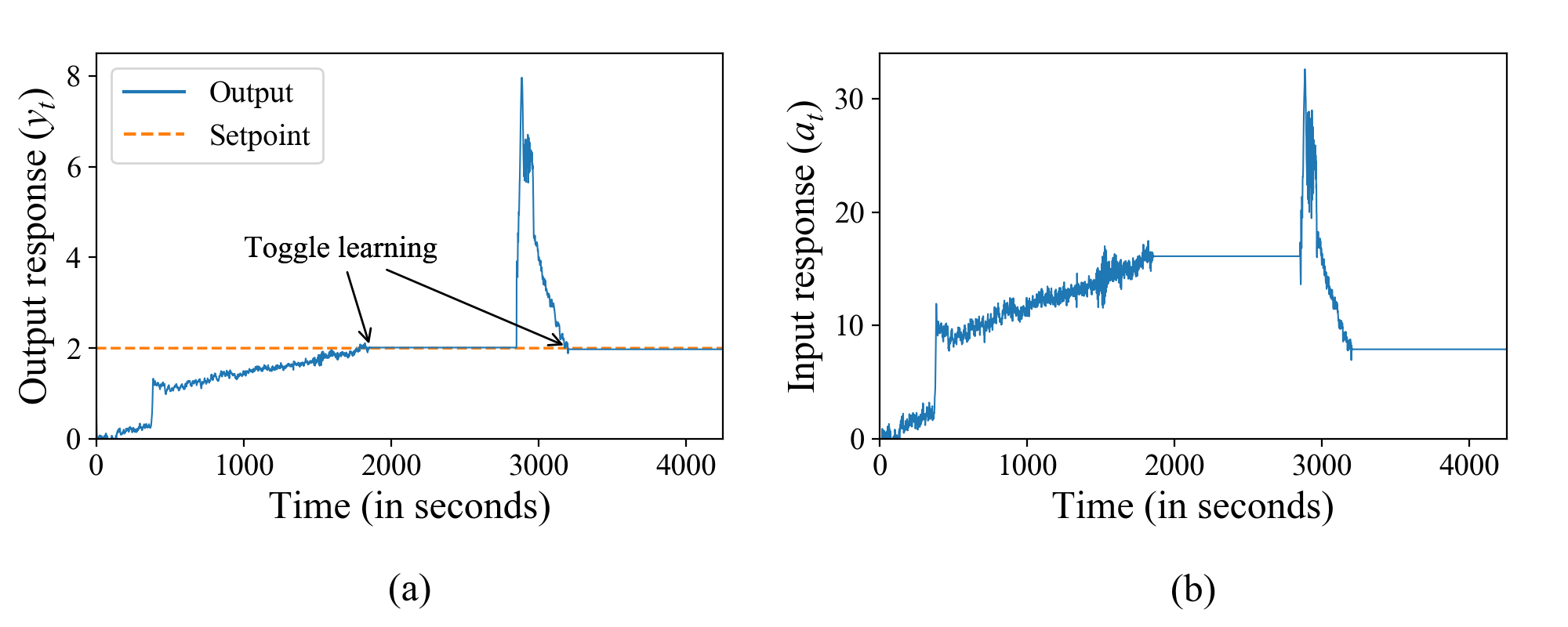}%
\caption{Simulation results for Example $4$ -- (a) set-point 
tracking performance of the DRL controller under abrupt process changes; (b) 
the controller action taken by the DRL controller.}
\label{fig:Example4}
\end{figure*}

We run Algorithm \ref{alg:ActorCriticProposed} continuously with the same specifications as in Example \ref{sec:Example1} for a fixed set-point $y_{sp}=2$. As shown in Figures \ref{fig:Example4}(a) and (b), in the approximate interval $0\leq t\leq 
1900$, the DRL agent is learning the control policy. Then at around $t = 1900$ we turn off the exploration noise and stop training the actor and critic networks because the moving average of the errors $|y_{t} - 2|$ across the past four time steps was sufficiently small (less than $0.0001$ in our case). Next, at around $t=2900$, a sudden process change is introduced by doubling the 
process gain in (\ref{eq:siso2}).
Observe that as soon as the process change is introduced, the policy learned by 
the DRL controller is no longer valid. Consequently, the system starts to deviate from the 
set-point. However, since the DRL controller learns in real-time, as soon as the process 
change is introduced, the controller starts re-learning a new policy.
Observe that for the same set-point $y_{sp}=2$, the DRL controller took about 
$400$ seconds to learn the new policy after the model change was introduced 
(see Figures \ref{fig:Example4}(a) and (b) for $t \geq 2900$).
This demonstrates how the DRL controller first learns the process 
dynamics and then learns the control policy -- both in real-time, without 
access to a priori process information.

\section{Comparison with Model Predictive Control}
\label{sec:DRLVsMPC}

Model Predictive Control (MPC) is well-established and widely deployed.
The proposed DRL controller has a number of noteworthy differences.

\renewcommand{\labelenumi}{(\alph{enumi})}
\begin{enumerate}[wide=0pt]

\item\textbf{Process model:}
An accurate process model is a key ingredient in any model predictive controller.
MPC relies on model-based predictions and model-based optimization
to compute the control action at each sampling time \cite{Steven2018}.
In contrast, the DRL controller does not require any \emph{a priori}
access to a process model: it develops the control policy in real time
by interacting with the process.

\item\textbf{Constraints:}
Industrial control systems are subject to operational and environmental
constraints that need to be satisfied at all sampling times.
MPC handles these constraints by explicitly incorporating them into
the optimization problem solved at each step \cite{wallace2016offset, lee2010approximate}.
In the DRL controller, constraints can be embedded directly in the reward function
or enforced through gradient-clipping---see (\ref{eq:Gradients}).
The latter approach is softer: constraint violation may occur during training,
but it is discouraged by triggering an enormous penalty in the reward signal.
Robust mechanisms for safe operation of a fully trained system, and indeed, for
safe operation during online training,
are high priorities for further investigation.

\item\textbf{Prediction horizon:}
MPC refers to a user-specified \emph{prediction horizon} when optimizing
a process's future response.
The duration of the MPC's planning horizon determines
the influence of future states; it may be either finite or infinite,
depending on the application.
The corresponding adjustment in the DRL controller comes through the discount
factor $\gamma\in[0,1]$ used to determine the present value of future rewards.

\item\textbf{State estimation:}
The performance of MPC relies on an accurate estimate of
the hidden process states over the length of the horizon.
The state estimation problem is typically formulated as a filtering problem \cite{tulsyan2016state, tulsyan2013particle},
with Kalman filtering providing the standard tool for linear processes,
while nonlinear systems remain an active area of research \cite{tulsyan2016particle, tulsyan2014performance}%
.
In this paper, we consider the DRL controller for cases where the full
system state is available for policy improvement.
Extending the design to include filtering and/or observer will be the subject
of future work.

\item\textbf{Adaptation:}
Practical control systems change over time.
Maintaining acceptable performance requires responding to both changes
and unknown disturbances.
Traditional MPC-based systems include mechanisms for detecting model-plant mismatch
and responding as appropriate $\ldots$ typically by initiating interventions in the
process to permit re-identification of the process model.
This process, which calls for simultaneous state and parameter estimation,
can be both challenging and expensive \cite{tulsyan2018switching, tulsyan2013simultaneous, tulsyan2013bayesianB}.
Some recent variants of MPC, such as robust MPC and stochastic MPC,
take model uncertainties into account by embedding them directly
into the optimization problem \cite{lee2010approximate}.
The DRL controller, on the other hand, updates the parameters
in the actor and critic networks at each sampling time using the latest experiences.
This gives the DRL controller a self-tuning aspect, so that process
remains optimal with respect to the selected reward criterion.

%
\section{Limitations}
 \label{sec:Limitations}
In the decades since MPC was first proposed its theoretical foundations
have become well established.
Strong convergence and stability proofs are available for several
MPC formulations, covering both linear and nonlinear systems \cite{mayne2014model}.
At this comparatively early stage,
the DRL controller comes with no such optimality guarantees.
Indeed, the assumptions of discrete-time dynamics and full state observation
are both obvious challenges demanding further investigation and development.
(For example,
while the temporal difference error for a discrete-time system
can be computed in a model-free environment, the TD error formulation
for a continuous-time system requires complete knowledge
of its dynamics \cite{bhasin2011reinforcement}.)
\textcolor{black}{Other issues of interest include the role of data requirements, computational complexity, non-convexity, over- and under-fitting,
performance improvement, hyper-parameter initialization.}

\end{enumerate}

\section{Conclusions}
We have developed an adaptive deep reinforcement learning (DRL) controller
for set-point tracking problems in discrete-time nonlinear processes.
The DRL controller is a data-based controller based on
a combination of reinforcement learning and deep-learning.
As a model-free controller, this DRL controller learns the control policy
in real time, simply by interacting with the process.
The efficacy of the DRL controller has been demonstrated
in set-point tracking problems on a SISO, a MIMO and a non-linear system
with external disturbances.
The DRL controller shows significant promise as an alternative to
traditional model-based industrial controllers, even though some
practical and theoretical challenges remain to be overcome. 
%
%
As both computing power and data volumes continue to increase,
DRL controllers have the potential to become an important tool in
process control.
\color{black}

\appendix
\section*{Basics of Neural Networks}
\label{sec:Basics_NN}

An artificial neural network (NN) is a mapping
$F\colon\mathcal{U}\rightarrow \mathcal{Y}$
from the input space $\mathcal{U}\subseteq\mathbb{R}^{n_u}$ to the
output space $\mathcal{Y}\subseteq\mathbb{R}^{n_y}$
whose internal structure is inspired by natural networks of neurons,
connected in several ``layers''.
Each layer defines a simple input-output relation of the form
\begin{align}
\label{eq:1-layer}
y=f(W u),
\end{align}
where  $W\in\mathbb{R}^{n_y\times n_u}$ is a weight matrix
carrying the vector space of inputs to the vector space of outputs,
and $f$ is a (typically nonlinear) ``activation function'' acting
elementwise on its input.
Common activation functions include
\begin{align*}
f(u)&=\sigma(u) := \frac{1}{1+\exp^{-u}},\\
f(u)&=\tanh(u) := \frac{\exp^{2u}-1}{\exp^{2u}+1},\\
f(u)&= ReLU(u) := \max\left\{0,u\right\}.
\end{align*}
The weight matrices and activation functions may differ from layer to layer,
so that a 3-layer NN model could have the general form
\begin{align}
\label{eq:3-layer}
y=W_{4}f_3(W_3 (f_2( W_2 (f_1( W_1 u))))).
\end{align}
Different sparsity patterns in the weight matrices $W_k$ allow for different
connections between neurons in successive layers.
Completing the definition of $F$ requires selecting the elements of the
weight matrices $W_k$: an optimization problem is formulated to minimize
the discrepancy between the input-output pairs produced by the mapping $F$
and a collection of input-output pairs the network is expected to ``learn''.
Typically the dimensions $n_y$ and $n_u$ are very large, so the number of
choice variables in this optimization is enormous.
Some form of gradient descent is the only practical scheme for the iterative
solution of the problem, a process evocatively called training the network.

\section*{Acknowledgments}
We would like to thank MITACS for financial support through the Globalink 
Graduate Fellowship.
We would also like to thank Kevin and Zilun for helpful discussions.



\section*{References}
\bibliographystyle{vancouver-authoryear}
\bibliography{RL_BIB.bib}

\end{document}